\documentclass[journal]{IEEEtran}

\IEEEoverridecommandlockouts
\usepackage{cite}
\usepackage{amsmath,amssymb,amsfonts}
\usepackage{algorithmic}
\usepackage{graphicx}
\usepackage{textcomp}
\usepackage{subcaption}
\usepackage{booktabs,caption}
\usepackage{threeparttable}
\usepackage{float}
\usepackage[table,xcdraw]{xcolor}

\def\BibTeX{{\rm B\kern-.05em{\sc i\kern-.025em b}\kern-.08em
    T\kern-.1667em\lower.7ex\hbox{E}\kern-.125emX}}

\usepackage[colorlinks=true, allcolors=blue]{hyperref}
\newcommand{\spBwFig}[1]{\vspace{-0.5cm}}

\begin{document}

\title{Below 100 ps CTR using FastIC+, an ASIC including on-chip digitization for ToF-PET and beyond}
% \title{FastIC+: A sub-90 ps ASIC including digitization on-chip for radiation detectors in ToF-PET}
% FastIC+: A sub-100 ps ASIC including digitization on-chip for radiation detectors in ToF-PET
% Author antic: FastIC+: An Analog Front-End including on-chip TDCs for fast timing detectors

\author{  D.~Mazzanti\href{https://orcid.org/0009-0003-9319-777X}        {\includegraphics[scale=0.08]{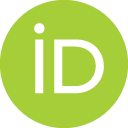}}, 
         S.~Gómez\href{https://orcid.org/0000-0002-3064-9834}{\includegraphics[scale=0.08]{Figures/orcid}}, 
         J.~Mauricio\href{https://orcid.org/0000-0002-9331-1363}{\includegraphics[scale=0.08]{Figures/orcid}},
         J.~Alozy,
         F.~Bandi,
         M.~Campbell,
         R. Dolenec,
         G.~El Fakhri,
         J.~M.~Fernández-Tenllado\href{https://orcid.org/0000-0003-1762-068X}{\includegraphics[scale=0.08]{Figures/orcid}},
         A.~Gola, 
         D.~Guberman\href{https://orcid.org/0000-0002-9636-1825}{\includegraphics[scale=0.08]{Figures/orcid}}, S.~Majewski\href{https://orcid.org/0000-0002-9645-7966}{\includegraphics[scale=0.08]{Figures/orcid}},
         R.~Manera\href{https://orcid.org/0000-0003-4981-6847}{\includegraphics[scale=0.08]{Figures/orcid}}, A.~Mariscal-Castilla\href{ https://orcid.org/0000-0002-7127-2535}{\includegraphics[scale=0.08]{Figures/orcid}},
         M.~Penna, 
         R.~Pestotnik\href{https://orcid.org/0000-0003-1804-9470}{\includegraphics[scale=0.08]{Figures/orcid}}, S.~Portero\href{https://orcid.org/0009-0001-6167-2500}{\includegraphics[scale=0.08]{Figures/orcid}},
         A. Paterno\href{https://orcid.org/0000-0002-2263-8932}{\includegraphics[scale=0.08]{Figures/orcid}},
         A.~Sanuy \href{ https://orcid.org/0000-0002-5767-3623}{\includegraphics[scale=0.08]{Figures/orcid}},
         J.~J.~Silva,
         R.~Ballabriga\href{https://orcid.org/0000-0002-7372-6131}{\includegraphics[scale=0.08]{Figures/orcid}},
         and 
         D.~Gascón\href{https://orcid.org/0000-0001-9607-6154}{\includegraphics[scale=0.08]{Figures/orcid}}   
\\	  

\thanks{
    This work did not involve human subjects or animals in its research.
    
	Authors D.~Mazzanti, A.~Mariscal-Castilla, R.~Manera, J.~Mauricio, S.~Portero, J.J.~Silva, A.~Sanuy, J.M.~Fernández-Tenllado, D.~Guberman and D.~Gascón work at the Dept. Física Quàntica i Astrofísica, Institut de Ciències Del Cosmos (ICCUB), Universidad de Barcelona (IEEC-UB), Barcelona, Spain (e-mail: dmazzanti@icc.ub.edu ). 
 
	Author S.~Gómez works as a Serra Hunter fellow at Polytechnic University of Catalonia (UPC), Barcelona, Spain,(e-mail: sergio.gomez-fernandez@upc.edu).
	
    Authors J.~Alozy, R.~Ballabriga, F.~Bandi, M.~Campbell, A.~Paterno work at Conseil Européen pour la recherche nucléaire (CERN), Meyrin, Switzerland
    
    Author G.~El Fakhri works at Yale PET Center, Yale University School of Medicine, New Haven, Connecticut, USA.

    Authors A.~Gola and M.~Penna work at Custom Radiation Sensors (CRS), Fondazione Bruno Kessler, Trento, Italy.

    Author S.~Majewski works at the Dept. of Biomedical Engineering, Univeristy of California, Davis, USA

    Author R.~Dolenec and R.~Pestotnik work at Experimental Particle Physics Department (F9), Jožef Stefan Institute, Ljubljana, Slovenia.
 
    D.~Mazzanti and S.~Gómez are co-first authors and corresponding authors.

}

\thanks{Manuscript received W, 2023; accepted X. Date of publication Y; date of current version Z.}
}

\maketitle

\IEEEpeerreviewmaketitle
	
	% In less than 2000 characters (including blanks, excluding headline authors, affiliations and keywords), summarize your findings, and describe the implications of those findings. The abstract must be an accurate reflection of what is in your article as follows:

\begin{abstract}
This work presents the 8-channel FastIC+, a low-power consumption and highly configurable multi-channel front-end ASIC with internal digitization, for the readout of photo-sensors with picosecond time resolution and intrinsic gain. This ASIC, manufactured in 65 nm CMOS technology, can readout positive or negative polarity sensors and provides a digitized measurement of the arrival time and energy of the detected events with a power consumption of 12.5 mW per channel. On-chip digitization is executed by a Time-to-Digital Converter (TDC) based on a Phase-Locked Loop (PLL) generating 16 phases at 1.28 GHz. The internal TDC introduces a jitter contribution of 31.3 ps FWHM, with minimal impact on timing measurements. When evaluating FastIC+ to readout 3$\times$3 mm$^2$ silicon photomultipliers (SiPMs) with a pulsed laser, we achieved a single-photon time resolution (SPTR) of (98 $\pm$ 1) ps FWHM. We also performed time-of-flight positron emission tomography (ToF-PET) experiments using scintillator crystals of different sizes and materials. With LYSO:Ce,Ca crystals of 2.8$\times$2.8$\times$20 mm$^3$ we obtained a coincidence time resolution (CTR) of (130 $\pm$ 1) ps FWHM. With LGSO crystals of 2$\times$2$\times$3 mm$^3$, a CTR of (85 $\pm$ 1) ps FWHM. To the best of our knowledge, this is the first time that a CTR below 100 ps using on-chip digitization is reported.

% This work presents the 8-channel FastIC+, a low-power consumption and highly configurable multi-channel front-end ASIC with internal digitization, for the readout of photo-sensors with picosecond time resolution and intrinsic gain. This ASIC, manufactured in 65 nm CMOS technology, can readout positive or negative polarity sensors and provides a digitized measurement of the arrival time and energy of the detected events with a power consumption of 12.5 mW per channel. On-chip digitization is executed by a Time-to-Digital Converter (TDC) based on a Phase-Locked Loop (PLL) generating 16 phases at 1.28 GHz. The internal TDC introduces a jitter contribution of 31.3 ps FWHM, with minimal impact on timing measurements. When evaluating the FastIC+ TDC on SPTR for an FBK NUV-HD-TM LFv2 M0 3$\times$3 mm$^2$ SiPM, we achieved 98 $\pm$ 1 ps FWHM. For CTR measurements, a sub-90 ps time resolution result of 85 $\pm$ 1 ps FWHM is achieved when coupling 2$\times$2$\times$3 Fast-LGSO crystals to the same SiPMs and 130 $\pm$ 1 ps when coupling 2.8$\times$2.8$\times$20 mm$^{3}$ LYSO:Ce,Ca crystals to FBK NUV-HD-MT LF M0 4$\times$4 SiPMs. These results demonstrate that FastIC+ can enhance the time resolution of current TOF-PET systems while maintaining low power consumption.

\end{abstract}

	\section{Introduction} \label{sec:Intro}

Time-of-flight Positron Emission Tomography (ToF-PET) has become an essential tool in medical imaging over the last years thanks to the improvement on the spatial resolution and the Signal-to-Noise Ratio (SNR) of PET scans compared to conventional PET systems \cite{Lecoq2017}. One of the key parameters that influences the performance of these scans is the Coincidence Time Resolution (CTR), which determines the accuracy of the time measurement of two detected 511 keV photons with antiparallel momenta resulting from a positron-electron annihilation. This is used to estimate the position of the annihilation along the so-called Line-of-Response. Therefore, improving the CTR is critical to improve the image resolution \cite{VANDENBERGHE201869,Schaart2021}, specially for whole-body PET applications \cite{CONTI20091, SiemensUSA}. 

Several factors influence the CTR performance, including the scintillation crystal, the photosensor and the readout electronics. Typically, fast scintillators based on Lutetium Oxyorthosilicate (LSO) and Lutetium Yttrium Orthosilicate (LYSO) have been used in ToF-PET detectors because of their excellent timing response \cite{Gundacker2020a}, high stopping power and fast-decay time ($\sim$ 40 ps). Regarding the photosensors, Silicon Photomultipliers (SiPMs) have become the detector of choice due to their compact size, high Photon Detection Efficiency (PDE) and insensitivity to magnetic fields \cite{Gonzalez-Montoro2022AInformation}. For instance, Fondazione Bruno Kessler (FBK) has developed near-ultraviolet SiPMs with a PDE close to 70$\%$ at 420 nm with reduced dark noise and crosstalk probability. Furthermore, an intrinsic Single Photon Time Resolution (SPTR) below 50 ps for a 3$\times$3 mm$^2$ SiPM has been achieved with this technology \cite{Gundacker_2023}. 

However, improving the time resolution requires not only optimizing scintillator and photodetector properties, but also the front-end electronics \cite{LECOQ8049484, GUNDACKER20156, Seifert_2012}. High-frequency (HF) amplifiers with low electronic jitter, (i.e., a large slew rate compared to the electronic noise) have been employed to study the fundamental limits of the photodetector-crystal pair \cite{Gundacker2019}. This solution has reported so-far the best timing performance \cite{Gundacker2020a, Nadig_2023}. However, the implementation of this readout circuit has been limited to small-scale prototypes due to their high power consumption (up to 150 mW/channel). In \cite{Cates2022}, the trade-off between power consumption and measured CTR has been studied for different amplifiers, showing that lower-power amplifiers can also achieve good timing results. Besides power consumption, this solution becomes impractical or suboptimal for scalability due to the physical dimensions of the electronics in large systems, where thousands of detectors are involved, and a more compact design is needed.

The digitization of timing and energy information is another important aspect at the system level. The HF readout only includes analog signal processing, requiring an additional circuit to digitize the data. Analog-to-Digital Converters (ADCs) are a suitable option, especially for energy measurements, where timing is not an issue \cite{Sanmukh2024}. Regarding timing measurements, a Time-to-Digital Converter (TDC) is a more power-efficient circuit for obtaining precise timestamps with low power consumption. Two trends in TDC implementation already exist \cite{MATRIX16}: (1) Field Programmable Gate Array (FPGA) based TDCs that use the fastest delay elements in the device which can achieve good timing with moderate power consumption, and (2) Application Specific Integrated Circuits (ASIC) based TDCs that can be customized for specific applications, offering better overall performance particularly concerning the power consumption, and giving a similar timing resolution. Additionally, ASIC-based TDCs are more compact and scalable since the TDC can be integrated into the same ASIC as the analog readout.

Over the years, several ASICs have been proposed as compact and power-efficient readout circuits. ASICs with analog outputs include the NINO ASIC \cite{NINO}, which provides only timing information, and the Weeroc RADIOROC2 \cite{saleem23}, which provides timing output as a binary signal and energy as an analog signal. Other ASICs that provide binary outputs include HRFlexToT and FastIC \cite{AntonioFastIC2024, SergioFastIC2021JINST, SergioFastIC2021}. Among the ASICs with fully digital outputs (integrating internal ADCs or TDCs) are the Weeroc PETIROC2A \cite{Petiroc2A} and the PETsys TOFPET series (TOFPET2, TOFPET3) \cite{TOFPET2_Nadig}.

In this work, we present the FastIC+, a low-power, multi-channel front-end ASIC with an integrated TDC. We describe its architecture, characterize the TDC performance using electric test signals, and evaluate the ASIC’s performance for the direct detection of optical photons using SiPMs, as well as for gamma-ray detection in PET applications using scintillators coupled to SiPMs.

	\section{FastIC+ Architecture} \label{sec:architecture}

The analog processing of the FastIC+ is similar to that used in the FastIC~\cite{FastIC_PMB,GomezFASTICJINST2022}. The digitization stage employs an event-driven TDC based on a Phase-Locked Loop (PLL) to achieve high-precision time tagging. Fig.~\ref{fig:Fastiplus_BlockDiagram} illustrates the architecture of the FastIC+, including both the analog and digital signal processing stages.

\subsection{Analog readout}
\begin{figure*}[htpb]	
	\begin{center}
		\includegraphics[width=\textwidth]{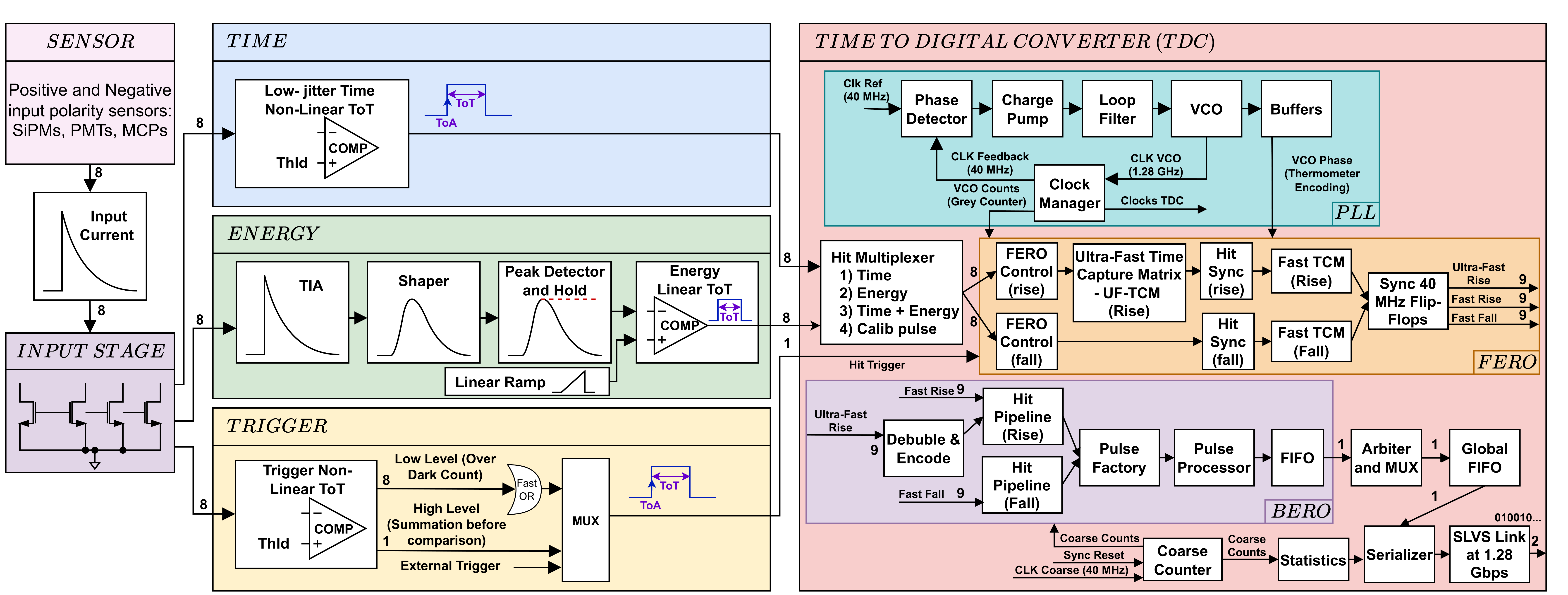}
		\vspace*{-0.2cm}
		\caption{FastIC+ block diagram, including both analog and digital signal processing.}  
		\vspace{-0.4cm}
		\label{fig:Fastiplus_BlockDiagram}
		\vspace{-0.25cm}
	\end{center}	
\end{figure*}

The signals from sensors with intrinsic amplification, such as SiPMs, multi-anode photomultiplier tubes, or microchannel plates (MCPs), are read out by a current-mode input stage. The input stage can deal with positive or negative polarity signals. A current conveyor, based on a double feedback mechanism~\cite{PaperHRFlexToT_TRPMS, PaperMUSIC_MDPI}, is employed to stabilize the input voltage and maintain a low input impedance within the desired bandwidth. Additionally, it creates three copies of the signals that are processed by three different branches: "Time", "Energy" and "Trigger", as shown in Fig.~\ref{fig:Fastiplus_BlockDiagram}.

The time path compares a copy of the signal delivered by the sensor to a threshold using a leading-edge current comparator with non-linear feedback to ensure high-speed operation~\cite{Traff1992NovelAT,Rio_Fernandez1997}. This path provides (1) precise photon Time-of-Arrival (ToA) information and (2) energy measurement using a low-jitter, non-linear Time-over-Threshold (ToT) response. 

The input stage trigger branch follows the same processing scheme but attenuates the input current, since its jitter requirements are less stringent, allowing for reduced power consumption. The trigger threshold can be adjusted independently of the time branch to validate events or serve as a second timing threshold for rise-time measurements~\cite{Kratochwil2021}. The trigger signal is generated as a logical OR across all channels, meaning that only the fastest channel’s timing information is available. Additionally, the time signal can serve as a trigger for low-light-level detection, and an external trigger can be used for calibration purposes. Lastly, a high-level trigger is generated by summing the input signals from all channels in the analog domain before the comparator stage.

The dedicated signal for energy measurement is initially converted from current to voltage using a Trans-Impedance Amplifier (TIA). This path also employs a shaper with a first-order active integrator to control the peaking time and a pole-zero cancellation network to reduce signal tailing. 

A peak-and-hold detector (PDH), based on~\cite{DEGERONIMO2002544}, captures the pulse amplitude. The PDH can be adjusted to function as a track-and-hold circuit and includes a discharge mechanism to prevent inaccuracies from undesired pulses (e.g., dark pulses below the trigger threshold), as described in~\cite{PaperHRFlexToT_TRPMS}. The detected peak is then compared to a linear ramp to emulate a single-slope ADC, producing a pulse whose duration is proportional to the input peak amplitude. The ramp starts with a lower DC voltage compared to the DC baseline of the energy signal to allow for the calibration of channel-to-channel variations. The energy pulse captured in the absence of an input signal is referred to as the pedestal and must be calibrated and then subtracted in the data analysis. The time and energy signals from all eight channels, along with the validation trigger, are then sent to the digitization stage.

\subsection{Phase-Locked Loop (PLL)}

Digitization requires low-jitter internal signals, achieved through a Phase-Locked Loop (PLL), which multiplies the 40 MHz reference clock by a factor of 32, resulting in a frequency of 1.28 GHz. The PLL comprises several key components: a Phase and Frequency Detector (PFD), a charge pump, a loop filter, a Voltage-Controlled Ring Oscillator (VCO) and a clock manager, which includes a frequency divider, and buffers.

The digital PFD compares the system reference clock at 40 MHz with a feedback clock generated from the VCO to determine whether the generated clock leads or lags the reference. This digital implementation allows control over how many cycles remain in the locked or unlocked state when detecting a phase difference, emulating hysteresis behavior. The charge pump charges and discharges the VCO’s control voltage based on the phase detector’s status. It is implemented using a regulated cascode current mirror with a programmable range to ensure an accurate current source. The VCO control voltage includes a buffer to mitigate charge-sharing effects. A loop filter with large capacitors is necessary to filter noise-induced fluctuations in the control voltage caused by the high VCO gain (7 GHz/V), ensuring a smooth control signal.

The VCO consists of 16 delay elements based on the current-steering differential delay cell~\cite{Maneatis1992}, connected in a ring, with one cell output cross-coupled to ensure an odd number of inversions. The impedance of the diode-connected transistors, which affects the circuit's RC constant and thus the delay, is regulated by the VCO's control voltage. This implementation provides a wide tuning range, excellent power supply rejection, and decouples the control voltage from the power supply while achieving a time bin (delay of the cell) of 24.4 ps.

A buffer is added to each of the VCO phases before sending them to the next stage, forming the 5 Least Significant Bits (LSBs) of the TDC. These buffers serve two purposes: isolating the VCO from the Time Capture Registers and enabling compensation for time bin non-uniformities (e.g., for calibration to improve time bin linearity). Each buffer includes configurable bits to adjust both the rising and falling edge delays. Additionally, the odd-numbered stages can be disabled to reduce power consumption, at the cost of increased time bin width (49~ps), as only half of the VCO phases are used in subsequent stages.

The clock manager consists of a frequency divider. It generates reference clocks for the TDC blocks derived from the VCO’s 1.28 GHz clock. These clocks include the coarse counter, the 40 MHz feedback clock for the phase detector, and the 40 MHz synchronization clock. It also generates VCO counts for the Front-End Readout (FERO) block, contributing to the 5 most significant bits (MSBs) of the TDC.

Simulations considering variations in process (fast, slow, and typical corners), temperature (from -40 ºC to 80 ºC), and voltage (from 1.08 V to 1.32 V, with a typical value of 1.2 V) show that the frequency of the VCO can be locked at 1.28 GHz by automatically adjusting the VCO's control voltage. The simulated jitter is below 5 ps, the phase noise is -80 dBc/Hz at 100 MHz, and the power consumption is 12 mW.

%We have 24ps time bin for the timing measurement and 390ps for the energy measurement.
% Finally, the TDC Back-End Readout processes the data, filters non-interesting events and transmits them via a serialized line which allows to transmit a sustained rate of up to 3 million pulses per second per channel

%Current-steering differential delay cell to (Maneatis cell) minimize both the circuit sensitivity to external interference noise and the injection of noise from the cell into the chip substrate. 

%The gate of the PMOS current source is now controlled to regulate the impedance of the diode-connected transistors that affects the RC constant at the circuit output and thus the delay.

% The design is Radiation Hard since this IP block will be reused in future Radiation Hard ASICs

% The PFD and the clock manager are designed using triple modular redundancy (TMR) \cite{TMRG} to shield the circuits against single event upsets (SEUs). However, the charge pump and the VCO are not triplicated, thus leaving the PLL vulnerable to single event transients that can result in brief periods of unlocked operation, lasting for microseconds. 

\subsection{Front-End Readout (FERO) block}

The digitization phase starts with the FERO block, which asynchronously receives binary pulses (hits) from the analog stage (time, energy, trigger, or both as consecutive pulses). The proposed architecture allows simultaneous rising and falling edge capture, as both are processed separately.

The FERO Control block, implemented using an Asynchronous Finite State Machine, ensures that multiple hits within the same 25 ns clock period are prevented. Pulses following an initial one are filtered to prevent the Time Capture Registers from firing twice, limiting the readout throughput to the 40 MHz reference clock. 

The Ultra-Fast Time Capture Matrix (UF-TCM) includes 16 sampling elements (master-slave flip-flops optimized for timing) that store the internal state of the PLL’s VCO at the rising edge of the hit, achieving a time bin of 24.4 ps. Timing resolution for the falling edge is less stringent, as it is used only for pulse width measurement. The hit signal is synchronized with the PLL’s 1.28 GHz clock frequency. The Fast Time Capture Matrix (TCM) uses standard-cell flip-flops to capture the VCO’s 5-bit Gray counter, which encodes the number of oscillations within a 25 ns period, for both rising and falling edges of the hit. The time bin of the TCM is 390 ps, corresponding to half the VCO clock period.

Finally, a synchronization block resamples the data at 40 MHz to mitigate metastability issues. From this stage onward, all data becomes synchronous, eliminating the need for stringent timing constraints. 

In total, one input hit generates 27 output bits for processing in subsequent stages. Rising edge information encodes 16 bits (thermometric encoding) from the UF-TCM, 5 bits from the Fast TCM. Falling edge information encodes 5 bits from the Fast TCM and 1 bit for the phase. The time bin for the rising and falling edges is 25 ps and 390 ps, respectively.

\subsection{The Back-End Readout (BERO) block}

The Back-End Readout (BERO) block processes the captured hit edges. The de-bubbling logic corrects TDC bubbles (errors in the digital output code that produce corrupted output codes after the sampling elements due to mismatches in delay elements, signal integrity issues or noise). In the FastIC+ architecture these may arise in the UF-TCMs, which sample VCO phases. After correction, the thermometer counter data is encoded into binary form to reduce the number of bits.

The Hit Pipeline (Rise, Fall) introduces a systematic latency of 75 ns to the acquired pulses by using three cascaded flip-flop stages operating at 40 MHz. This delay allows an external trigger system to perform event validation (such as a coincidence trigger in a PET system) and determine whether to store the already acquired data based on the validation signal. The external validation signal can be injected into the ASIC through the same external trigger input, while the energy acquisition trigger must rely on internal triggers.

The pulse factory combines rising and falling edges to construct a single data packet. Additionally, it can filter out pulses not validated by an external trigger or discard invalid pulses, such as those with consecutive rising or falling edges.

The Pulse Processor encodes the rising and falling edge counters into Timestamps (ToA) using the Ultra Fast, Fast, and Coarse Rise Counters, and Pulse Width (ToT) using the Fast and Coarse counters for both edges. It also provides the option to filter pulses by width. Note that the BERO block receives the coarse counter (12 bits), generated by a 40 MHz clock from the frequency divider. Finally, pulses are stored in the channel FIFO (First Input First Output) queue, awaiting the Arbiter's grant.

% This can be potentially dangerous in case that the error occurs in the MSB, which holds the VCO phase information.

% HAy que añadir el coarsse counter aqui. Preguntar a joan si añadirlo dentro del BERO para simplificar dibjo?

% AQUI: The coarse counter counts at 40 MHz and extends the dynamic range of both ToA and ToT
%ABAJO:  Moreover, it provides synchronization at the detector level with the Synchronous Reset pin

\subsection{Data transmission}

The Arbiter and Multiplexer receive bus requests from the TDC channels and manage data transmission according to two arbitration policies: Round-Robin (RR) and Pulse Sorting by Timestamp (PST). In the RR mode, events are transmitted cyclically in the channel order, while in PST mode, events are sorted based on their arrival time, facilitating efficient data post-processing.

The multiplexed data from the Arbiter is stored in a global FIFO queue before being serialized and transmitted via a single Scalable Low-Voltage Signaling (SLVS) output. The data transmission protocol used at the link layer is Aurora 64/66~\cite{Aurora}, supporting data rates ranging from 80 Mbps to 1.28 Gbps. This protocol enables the transmission of the information of up to 3 million pulses per second per channel (3 MHz/channel), ensuring high-speed data transfer.

Additionally, the TDC generates statistics, such as the number of filtered or discarded hits, at a low rate to aid in debugging and troubleshooting readout issues. To prevent overflow errors at low rates, the system also features an extended Coarse Counter (12 additional bits, for a total of 24 bits).

\subsection{FastIC+ Configurability}
\label{sec:fastic_configurability}

Several internal registers are included in the ASIC to configure various parameters. These registers can be modified via the I2C communication protocol. The ASIC can operate in \textit{analog mode}, functioning like its predecessor \textit{FastIC}, where each channel provides either an analog or SLVS response. Alternatively, in \textit{digital mode}, the Time-to-Digital Converter (TDC) is activated, and all information is transmitted through a single data link. 

\begin{figure*}[htp!]	
	\begin{center}
		\includegraphics[width=0.91\textwidth]{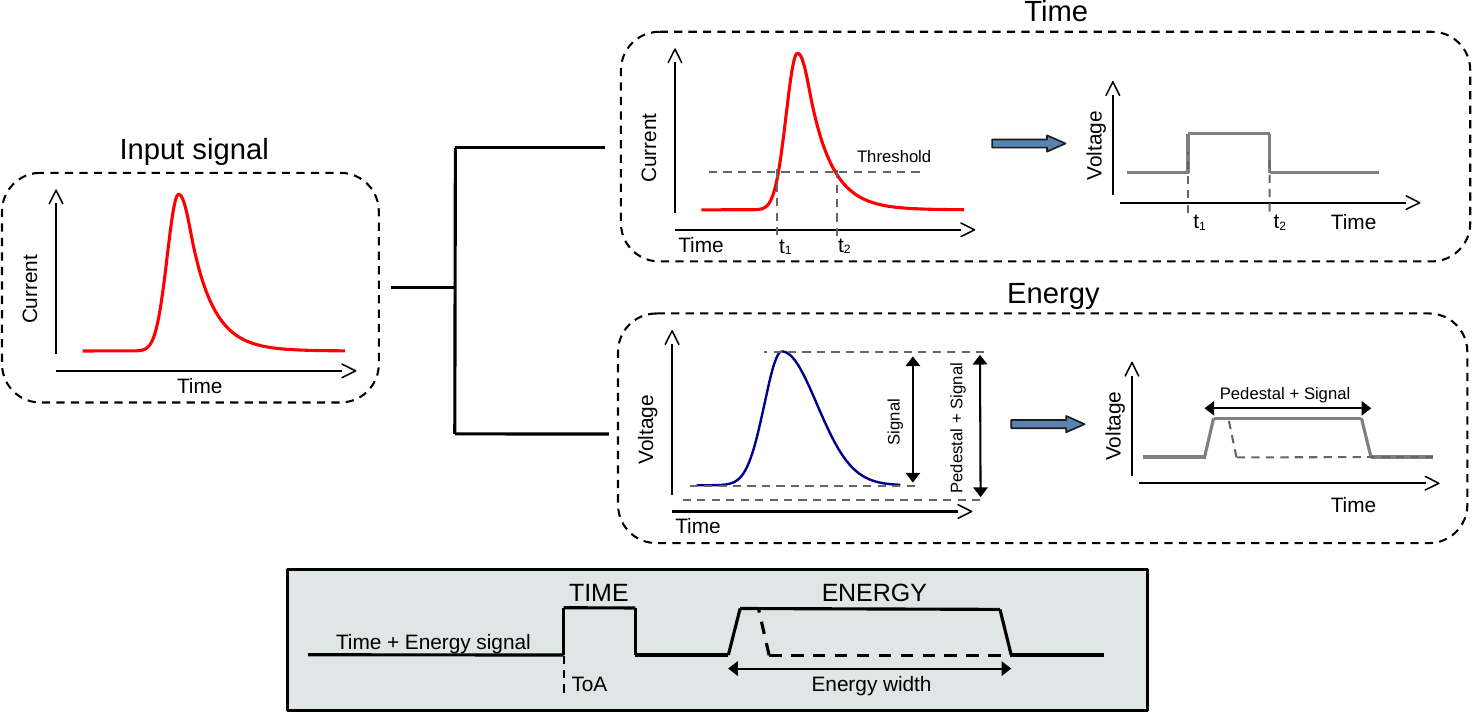}
		%\vspace*{-0.2cm}
		\caption{Response of the FastIC+ in High Energy Resolution mode, where Time (ToA) and Energy (ToT) are transmitted.}  
		%\vspace{-0.4cm}
		\label{fig:SignalGeneration}
		%\vspace{-0.25cm}
	\end{center}	
\end{figure*}

The TDC operates in four transmission modes, depending on the type of information encoded in the data link. In \textit{High Energy Resolution mode}, both the Time signal (Time-of-Arrival, ToA) and Energy (Time-over-Threshold, ToT) are transmitted. The falling edge of the time pulse is generated arbitrarily to reduce the pulse width, thereby increasing the maximum allowable event rate when the ToT of the Time signal is not required. In \textit{High-Speed mode}, only the ToA and the non-linear ToT of the Time signal are provided. \textit{Hybrid mode} combines the ToA and non-linear ToT of the Time signal with the standard ToT energy measurement. Finally, in \textit{Single Pulse mode}, either the ToA or the ToT of the Time signal is transmitted, depending on the configuration. Fig.~\ref{fig:SignalGeneration} illustrates how ToA and ToT (named as energy width) are generated from the time and energy paths, respectively, and transmitted in High Energy Resolution mode.

    \section{Materials and methods} \label{sec:Methods}

In this section we describe the SiPMs and scintillators employed and the setups used to asses the SPTR and CTR measurements. All measurements were conducted in a light-tight box under room temperature conditions ($\sim$ 20 $^{\circ}$C).

\subsection{Crystals and SiPMs}
The scintillators used are summarized in Table~\ref{tab:crystals}. All crystals were wrapped with at least 5 layers of Teflon and glued to the SiPMs using Cargille Melmount with a refractive index of $n_D$ = 1.582 at 588 nm and $\sim$ 25 $^{\circ}$C. The characteristics of the SiPMs are summarized in Table~\ref{tab:SiPMs}.

\begin{table*}[htp!]
    \centering
    \caption{Crystals employed in CTR measurements.}
    \begin{tabular}{|c|c|c|c|c|}
    \hline
    \rowcolor[HTML]{C0C0C0} 
    {\color[HTML]{000000} \textbf{Material}} & {\color[HTML]{000000} \textbf{Geometry (mm³)}} & {\color[HTML]{000000} \textbf{Decay time (ns)}} & {\color[HTML]{000000} \textbf{ILY$^a$ (ph. $\cdot$ keV$^{-1}$)}} & {\color[HTML]{000000} \textbf{Manufacturer}} \\ \hline
    \rowcolor[HTML]{E3EBAB} 
    LYSO:Ce,Ca$^b$ & \begin{tabular}[c]{@{}c@{}}2$\times$2$\times$3 \\ 2.8$\times$2.8$\times$20 \\ 3$\times$3$\times$20 \end{tabular} & 35 & 30.0 & Taiwan Applied Crystal \\ \hline
    \rowcolor[HTML]{D9EBF1}
    LYSO:Ce,Ca$^c$ & \begin{tabular}[c]{@{}c@{}}2$\times$2$\times$3 \\ 3.12$\times$3.12$\times$20\end{tabular} & 40 & 45.0 & Crystal Photonics Inc. \\ 
    \hline
    \rowcolor[HTML]{E3EBAB}
    LYSO:Ce$^d$ & 3$\times$3$\times$20 & 36 & 33.2 & Saint-Gobain \\ \hline
    \rowcolor[HTML]{D9EBF1}
    Fast-LGSO:Ce$^e$ & 2$\times$2$\times$3 & 33 & 90\% of a NaI:Tl ref. & Oxide Corporation \\ \hline
    \end{tabular}
    \label{tab:crystals}

    \begin{tablenotes}
    \centering
    \item $^a$ Intrinsic Light Yield.
    \item $^b$ From \cite{Nadig_2023} \\
    \item $^c$, $^d$ From \cite{AntonioFastIC2024} \\
    \item $^e$ Oxide Corporation datasheet: \url{https://www.opt-oxide.com/en/products/fast-lgso/} \\
    \end{tablenotes}
\end{table*}

\begin{table*}[htp!]
    \centering
    \caption{SiPMs used in SPTR and CTR measurements.}
    \begin{tabular}{|c|c|c|c|}
    \hline
    \rowcolor[HTML]{C0C0C0} 
    {\color[HTML]{000000} \textbf{Name}} & {\color[HTML]{000000} \textbf{Area (mm$^2$)}} & {\color[HTML]{000000} \textbf{SPAD size ($\mu m^2$)}} & {\color[HTML]{000000} \textbf{V$_{break.}$$^a$ (V)}} \\ \hline
    \rowcolor[HTML]{E3EBAB} 
    FBK NUV-HD-MT LF M0 & \begin{tabular}[c]{@{}c@{}} 4$\times$4 \\ 1$\times$1 \\ Single SPAD \end{tabular} & 40$\times$40 & 32.5 \\ \hline
    \rowcolor[HTML]{D9EBF1} 
    FBK NUV-HD-MT LFv2 M0 & 3$\times$3 & 50$\times$50 & 32.5 \\ \hline
    \rowcolor[HTML]{E3EBAB} 
    HPK S13360-3050VE & 3$\times$3 & 50$\times$50 & 51.8 \\ \hline
    \end{tabular}
    \label{tab:SiPMs}

    \begin{tablenotes}
        \centering
        \item $^a$ Breakdown voltage.
    \end{tablenotes}
\end{table*}

\subsection{FastIC+ PCB module}
\begin{figure}[htp!]
    \centering
    \includegraphics[width=1\columnwidth]{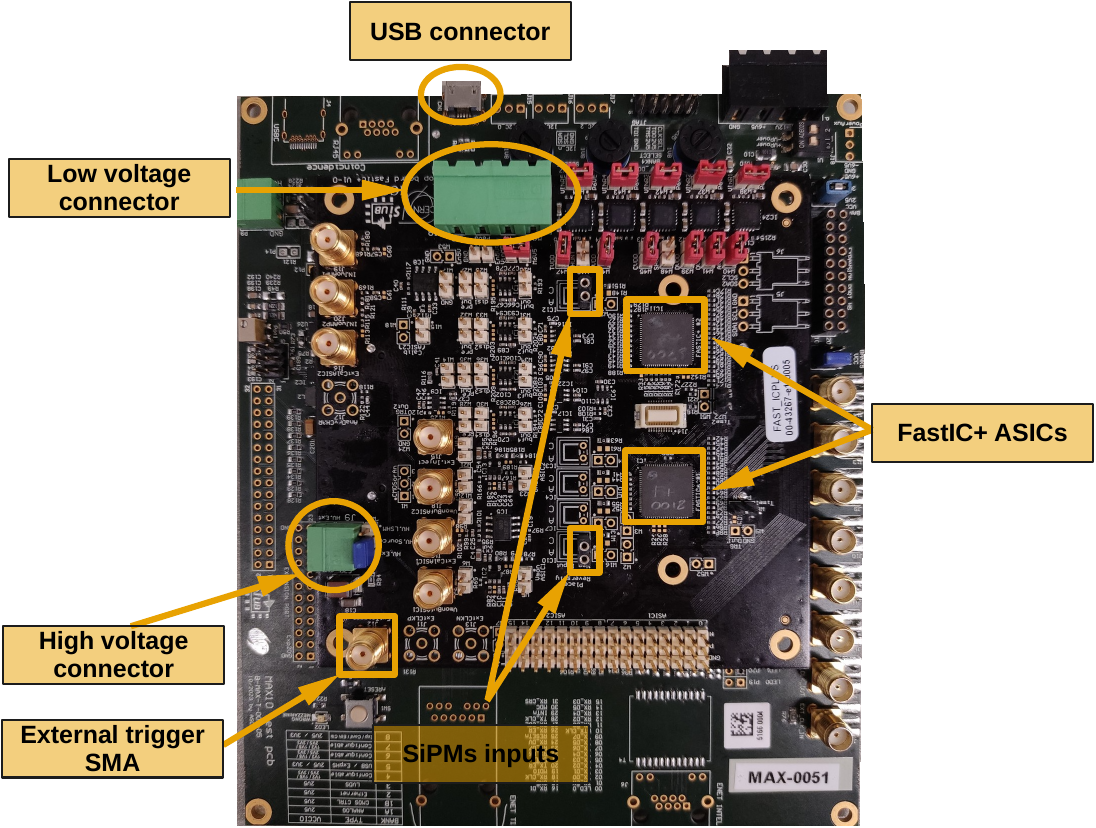}
    \caption{FastIC+ module designed to test the FastIC+ ASIC.}
    \label{fig:Module}
\end{figure}

We tested the FastIC+ ASIC using the FastIC+ evaluation module, as illustrated in Fig.~\ref{fig:Module}.
This module consists of two printed circuit boards (PCBs). The top board primarily houses two FastIC+ ASICs, allowing up to 16 SiPMs channels to be tested. It includes connectors to power both ASICs (which also incorporate voltage regulators to ensure a stable power supply) and the SiPMs. The TDC digitized output is sent to the bottom board, which hosts an FPGA. This FPGA transmits the acquired measurements to a computer for analysis via USB. Additionally, it manages the configuration of the ASICs via an I2C master.

\subsection{Electrical setup}

\begin{figure}[htp!]
    \centering
    \includegraphics[width=1\columnwidth]{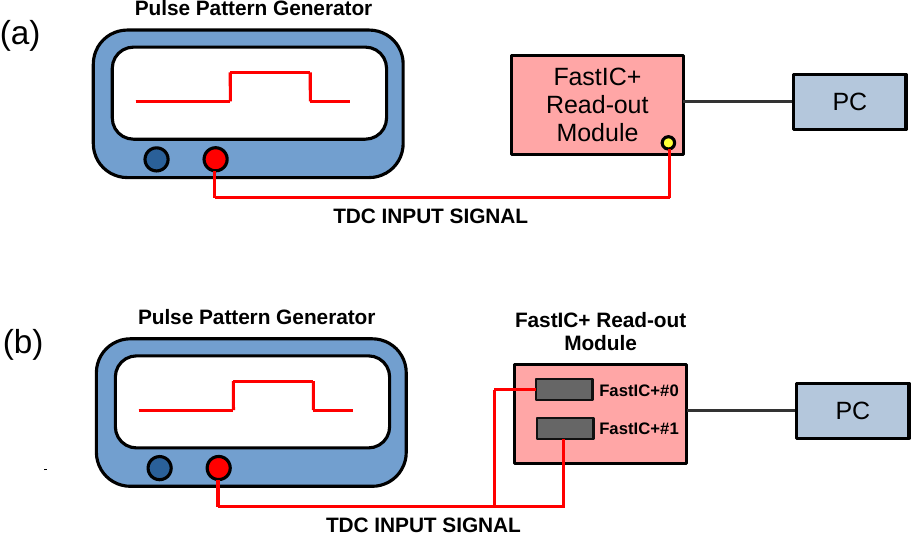}
    \caption{Schematic representation of the electrical setup employed to evaluate the FastIC+ TDC. (a) Code density test; (b) Single-shot precision of the TDC.}
    \label{fig:ElectricalTest}
\end{figure}

The evaluation of the TDC inside the FastIC+ has been carried out using the experimental setups shown in Fig.~\ref{fig:ElectricalTest}. The intrinsic linearity of the TDC is evaluated in terms of Differential Non-Linearity (DNL) and Integral Non-Linearity (INL) by performing a code density test using the setup shown in Fig.~\ref{fig:ElectricalTest} (a). The number of bins is computed using Eq.~\ref{eq:Nbins},

\begin{equation} \label{eq:Nbins}
N_{bins} = 2\cdot\frac{f_{osc_{PLL}}}{f_{clk_{ref}}} \cdot N_{cells} = 1024 \ ,
\end{equation}

\noindent where $f_{\text{osc}\text{PLL}}$ is the PLL oscillation frequency (1.28~GHz), $f_{\text{clk}_\text{ref}}$ is the reference clock frequency (40~MHz), $N_{\text{cells}}$ is the number of delay cells in the VCO (16 in this case), and the factor of 2 accounts for the two transitions (rising and falling edges) required for the VCO to complete one full oscillation cycle.

The procedure to compute the DNL using a density code test analysis, as described in \cite{DensityCode}, is as follows. The TDC captures one million pulses generated by an external Pulse Pattern Generator (Agilent 81110A), with random arrival times following a uniform distribution relative to the TDC’s reference clock. The number of detected pulses (counts) is categorized into time bins to construct a histogram. Ideally, this distribution should be flat, as the input signal follows a uniform distribution and the TDC is expected to assign events equally across all time bins. However, non-linearities within the TDC distort this ideal outcome, resulting in a non-uniform distribution. The deviation of the actual bin counts from the ideal uniform value is used to determine the DNL.

In particular, the normalized number of counts per time bin, i.e., the number of counts in a given time bin ($\text{counts}_{\text{bin}_i}$) divided by the average number of counts per time bin ($\text{counts}_{\text{average}}$), corresponds to the actual width of time bin $i$. Therefore, the DNL of time bin $i$ can be computed using Eq.~\ref{eq:DNL},

\begin{equation} \label{eq:DNL}
DNL_{bin_i} \ [ps] = \left(\frac{counts_{bin_i}}{counts_{average}}-1\right) \cdot TDC_{LSB} \ ,
\end{equation}

%DNL: Difference between an actual step width and the ideal step width.

\noindent where the ideal time bin size is 1, and the TDC's Least Significant Bit (LSB) corresponds to 24.4 ps. Finally, the Integral Non-Linearity (INL), which represents the cumulative deviation of the actual time bin widths (i.e., the integrated DNL), can be computed using Eq.~\ref{eq:INL}, 

\begin{equation} \label{eq:INL}
INL_{bin\_i} \ [ps] = \sum_{j=0}^{i-1} DNL_{bin\_j} \ .
\end{equation}

Fig.~\ref{fig:ElectricalTest}b illustrates the measurement setup used to compute the single-shot precision of the TDC, which accounts for intrinsic jitter, linearity deviations, and quantization error. In this setup, it is important to highlight that the TDC input signal is sent to two different ASICs, each with its own independent PLL, to obtain two completely independent measurements. An external trigger signal from the same Pulse Pattern Generator as before is sent to two channels from different ASICs. The ToA is computed as the time difference between the timestamps obtained from each ASIC. Consequently, the standard deviation of the jitter—representing the single-shot precision of a single TDC—is obtained using Eq.\ref{eq:jitter_TDC}, 

\begin{equation} \label{eq:jitter_TDC}
jitter_{TDC} \ [ps] = \frac{std\left(Time_{ASIC\_0}-Time_{ASIC\_1}\right)}{\sqrt{2}} \ .
\end{equation}

% Both TDC and PPG have different time references:
% If we build a histogram, we should expect a flat distribution where the normalized number of hits per time bin (i.e. number hits divided (by the average numner of hits per code)) correspond to the TDC bin width.
% A histogram with the #hits per time bin is built (total: 1024 bins) 
% Data is normalized (by the average numner of hits per code) to the CLK_REF period (25 ns).

\subsection{Single Photon Time Resolution} \label{sec:Methods-SPTR}
\begin{figure}[htp!]
    \includegraphics[width=\columnwidth]{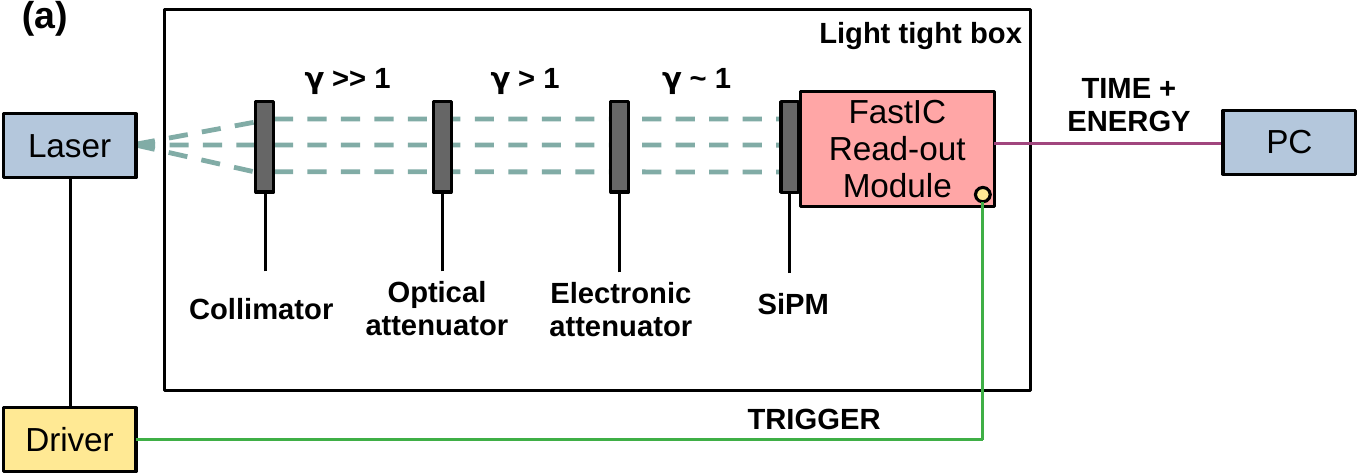} \\
    \vskip 1pt
    \includegraphics[width=\columnwidth]{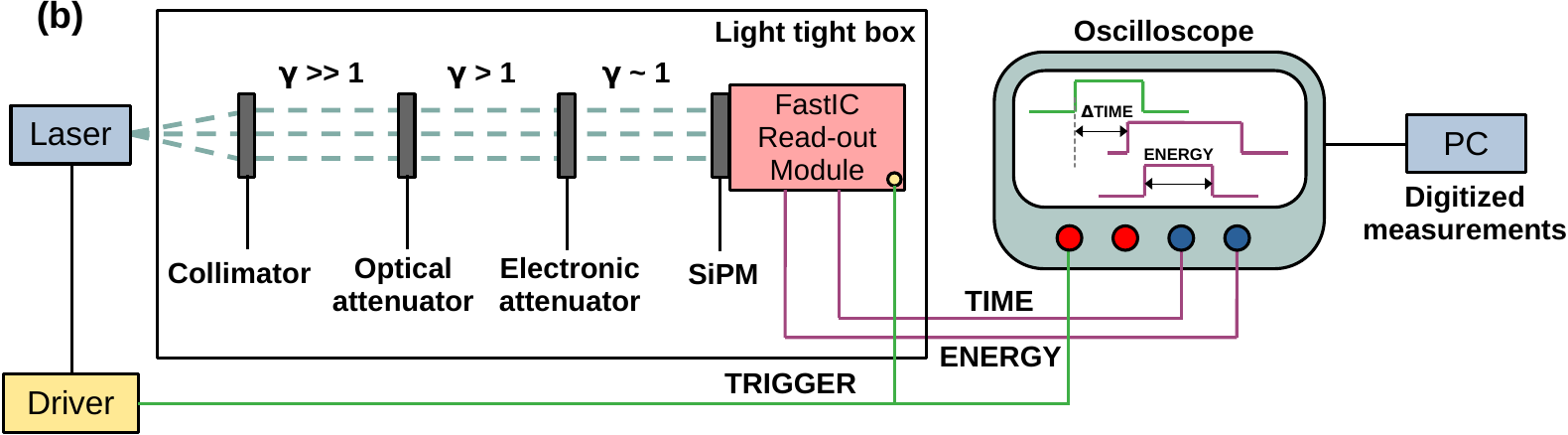}
    \caption{Schematic representation of the experimental setup for SPTR measurements. (a) Digital mode; (b) Analog mode.}
    \label{fig:SPTR-setup}
\end{figure}

Fig.~\ref{fig:SPTR-setup} provides a schematic representation of the experimental setups employed for the SPTR measurements. The light source used was a 405 nm pico-second laser (PiLas) with a pulse width of 28 ps FWHM. The laser operated at a repetition rate of 100 kHz and was tuned at a intensity level of 50 $\%$. Initially, the laser light passes through a collimator. Between the collimator and the SiPM, two different light attenuators were located to reduce the light intensity close to the single-photon level. First, a fixed optical attenuator decreased the light intensity down to a few photons. The second one was a liquid crystal attenuator (LCC1620/M from Thorlabs) in which the attenuation level can be modified by adjusting its operating voltage. Finally, the light reached the SiPM, which was connected to one channel of the FastIC+ ASIC module. The trigger signal generated by the laser driver was fed into the FastIC+ external trigger input.

To evaluate the contribution of FastIC+ TDC to the SPTR we performed the measurements both in analog (Fig.~\ref{fig:SPTR-setup}a) and digital (Fig.~\ref{fig:SPTR-setup}b) mode (see Sec.~\ref{sec:fastic_configurability}). In the analog mode, the FastIC+ binary signals (containing the information of the ToA and the energy width) along with the trigger signal provided by the laser driver, were sent to independent channels of an Agilent MS09254A oscilloscope (2.5 GHz, 20 GSa/s, 10 bits ADC). In the digital mode, the binary and trigger signals were processed and digitized by FastIC+. In both cases, the acquisition was triggered by the laser trigger signal.

For both measurements, the energy width was used for event selection. Therefore, a ToA distribution was constructed using the ToA values of these selected events. This distribution was fitted with a Gaussian function, and the time resolution was determined as the FWHM of the fit.

% For both measurements, the pulse width of the energy signal was used to select the events contributing to the time resolution. From these selected events, a ToA distribution was constructed using the timestamp of the time signal. The resulting distribution was then fitted to a Gaussian function, and the time resolution was determined as the FWHM of the fit, given by $FWHM = 2 \sqrt{2 \; \ln(2)} \; \sigma_{SPTR}$, where $\sigma_{SPTR}$ is the standard deviation.

\subsection{Coincidence Time Resolution}
\begin{figure}[h!]	
    \includegraphics[width=\columnwidth]{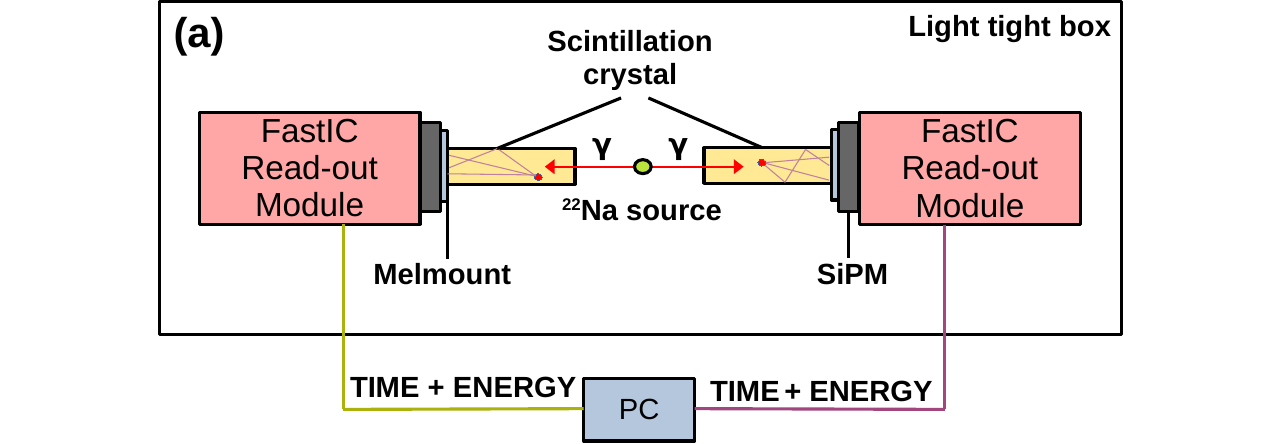}
    \vskip 10pt
    \includegraphics[width=\columnwidth]{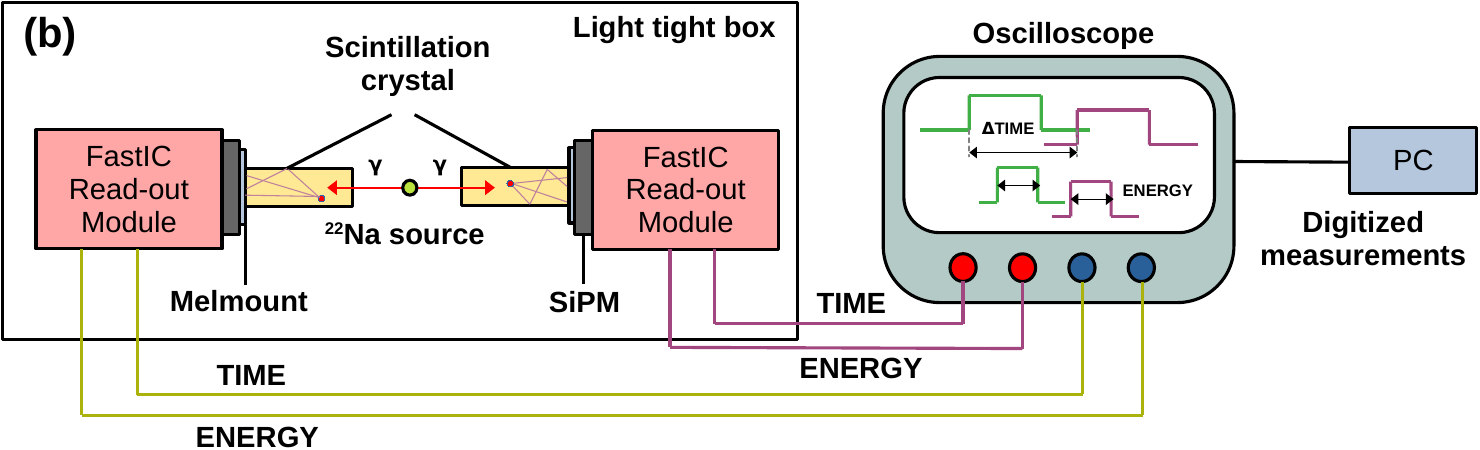}
    \caption{Schematic representation of the experimental setup for CTR measurements. (a) Digital mode; (b) Analog mode.}
    \label{fig:CTR-setup}
\end{figure}

Fig.~\ref{fig:CTR-setup} shows a schematic representation of the experimental setup used for the CTR measurements. A $^{22}$Na source was positioned and aligned between the two scintillator detectors. Moreover, each SiPM from each FastIC+ module was connected to an ASIC channel.

In this case, time and trigger comparators were used to acquire the events. The threshold for the time comparator was set to the single-SPAD signal level to achieve optimal timing performance. Conversely, the trigger comparator was set to its maximum value to reduce the number of undesired events caused by dark counts or low-energy interactions. The acquisition procedure was similar to those described in Sec.~\ref{sec:Methods-SPTR}, for both analog and digital signals. For the analog readout, a 25~ns coincidence time window was set on the oscilloscope between the two energy signals to ensure that both gamma photons originated from the same positron-electron annihilation, thus filtering out unwanted events. For digital measurements, an offline ToA condition was applied during analysis to classify events coming from different boards.

	\section{Performance evaluation} \label{sec:Res}
% This section provides a deep analysis of the TDC performance of the FastIC+. First, an electrical evaluation has been carry out to verify the specifications of the TDC. Second, the SPTR and CTR performance when using the analog and the digital readout is compared. Lastly, different scintillators are tested with the FastIC+.

\subsection{FastIC+ general performance}

The FastIC+ has been developed using a CMOS 65 nm technology node, with a die size of approximately 2.9$\times$2.9 mm$^2$ and housed in a standard QFN88 package. The TDC consumes 3.3 mW per channel, contributing to an overall power consumption of 12.5 mW per channel for the FastIC+. This represents an increase of only $\approx$ 0.5 mW/channel compared to FastIC. This small difference arises from the fact that the SLVS drivers per channel, which were previously used to send binary signals (time and energy) outside the chip, are no longer required in the FastIC+, as all data is transmitted through a single SLVS driver.

Additionally, a low-power mode has been implemented by reducing the number of buffers driving the VCO phases to the TDC, thereby lowering switching activity. In this mode, the TDC provides a time bin of approximately 49 ps (instead of 24.4 ps) while consuming just 1.8 mW per channel.

\begin{figure}[htpb!]	
    \includegraphics[width=\columnwidth]{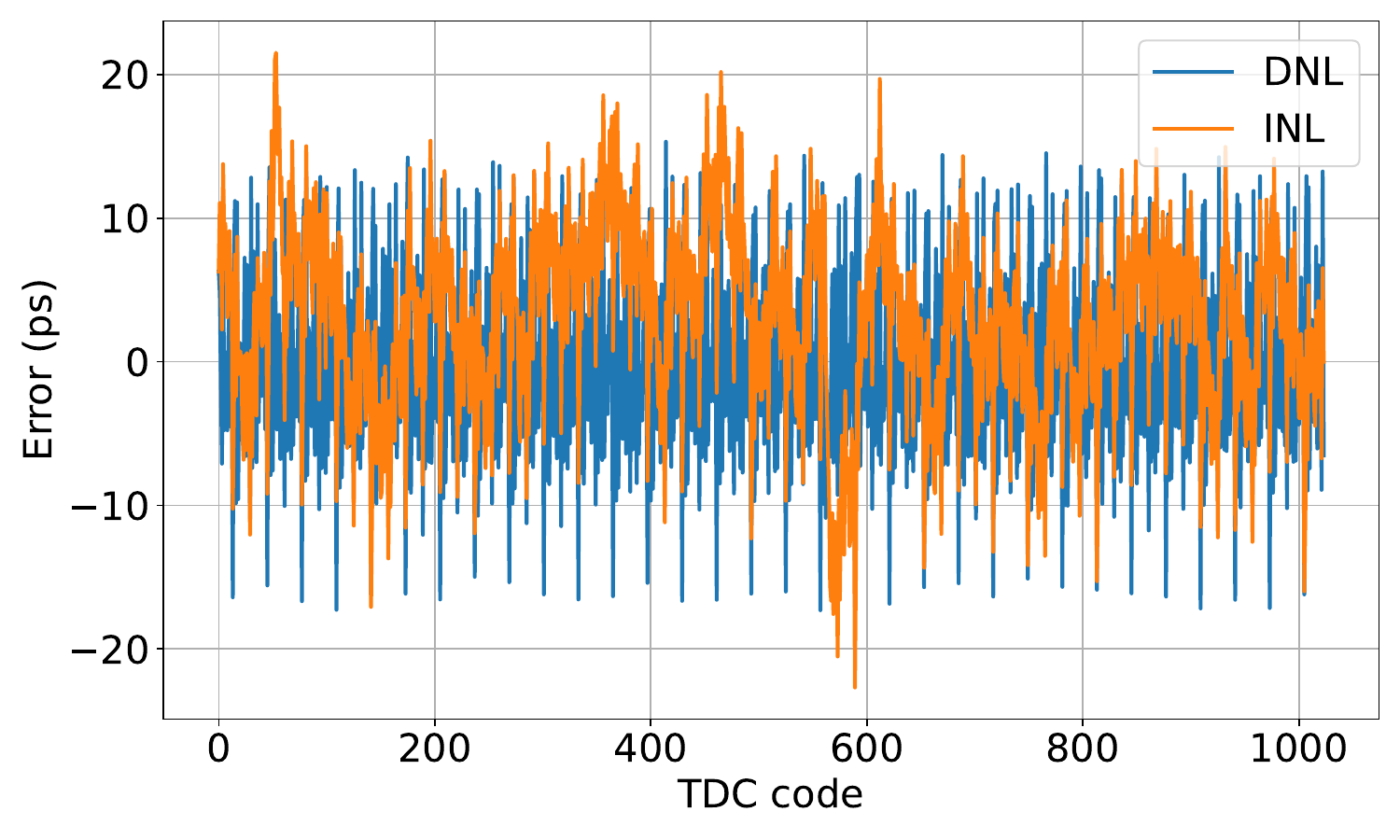}
    \caption{DNL and INL results after performing the code density test of 1 million pulses.}
    \label{fig:DNL_INL}
\end{figure}

Fig.~\ref{fig:DNL_INL} shows the DNL and INL results from a code density test. The standard deviation of the DNL is below 6~ps without calibration and improves to below 4~ps with time bin calibration, achieved by adjusting the drive strength of the PLL output buffers supplying the VCO phases. In both cases, the DNL remains lower than the LSB. The INL stays within $\pm$~22 ps. Finally, the single-shot precision is 13.3~ps (standard deviation) and 31.3~ps (FWHM).

%In April 2024, we received the FastIC+. Preliminary measurements of the packaged ASIC on a socket board injecting an electrical signal with a Pulse Pattern Generator (Agilent 81110A), revealed an Integral Non-linearity of $\pm$ 22 ps (lower than 1 LSB), a Differential Non-linearity standard deviation of 6.3 ps and an RMS jitter of 14.4 ps for the Time-to-Digital Converter. This results in a TDC resolution of 14.5 ps sigma or 34.1 ps FWHM ($\sqrt{\sigma_{DNL}^2 + \sigma_{rms\_jitter}^2}$). Incorporating the TDC contribution to the previous measurement, we expect that the SPTR will increase to \mbox{154.8 ps} and the CTR to \mbox{135.9 ps}, respectively. A low-power mode has been implemented by reducing the number of buffers that drive the VCO phases to the TDC, thereby reducing switching activity. In this mode, the FastIC+ will provide a time bin of about 49 ps (instead of 24 ps) with a power consumption of 1.8 mW/ch. 

% The DNL standard deviation is kept below the theoretical. That is because of the Jitter of the measurement, which dominates over the linearity.

%The jitter contribution of the TDC is approximately 22 ps FWHM, with only a $\sim$ 5$\%$ increase in power consumption per channel, having minimal impact on analog time resolution measurements. 

% 

% STD DNL (ps)  6.3
% RMS Jitter (ps) 14.4
% TDC Resolution (ps) 15.7

%XVDD + CVDD power (PLL & FERO): 22.8 mW
%OVDD power (TDCOUT Serializer only): 2.1 mW
%DVDD power (digital readout): 1.8 mW
%Total: 26,7 mW

\subsection{Digital vs analog performance for SPTR} \label{sec:Res-TDC_vs_Scope}
\begin{figure*}[htp!]
    \centering
    \begin{minipage}[t]{0.49\textwidth}
        \centering
        \includegraphics[width=\textwidth]{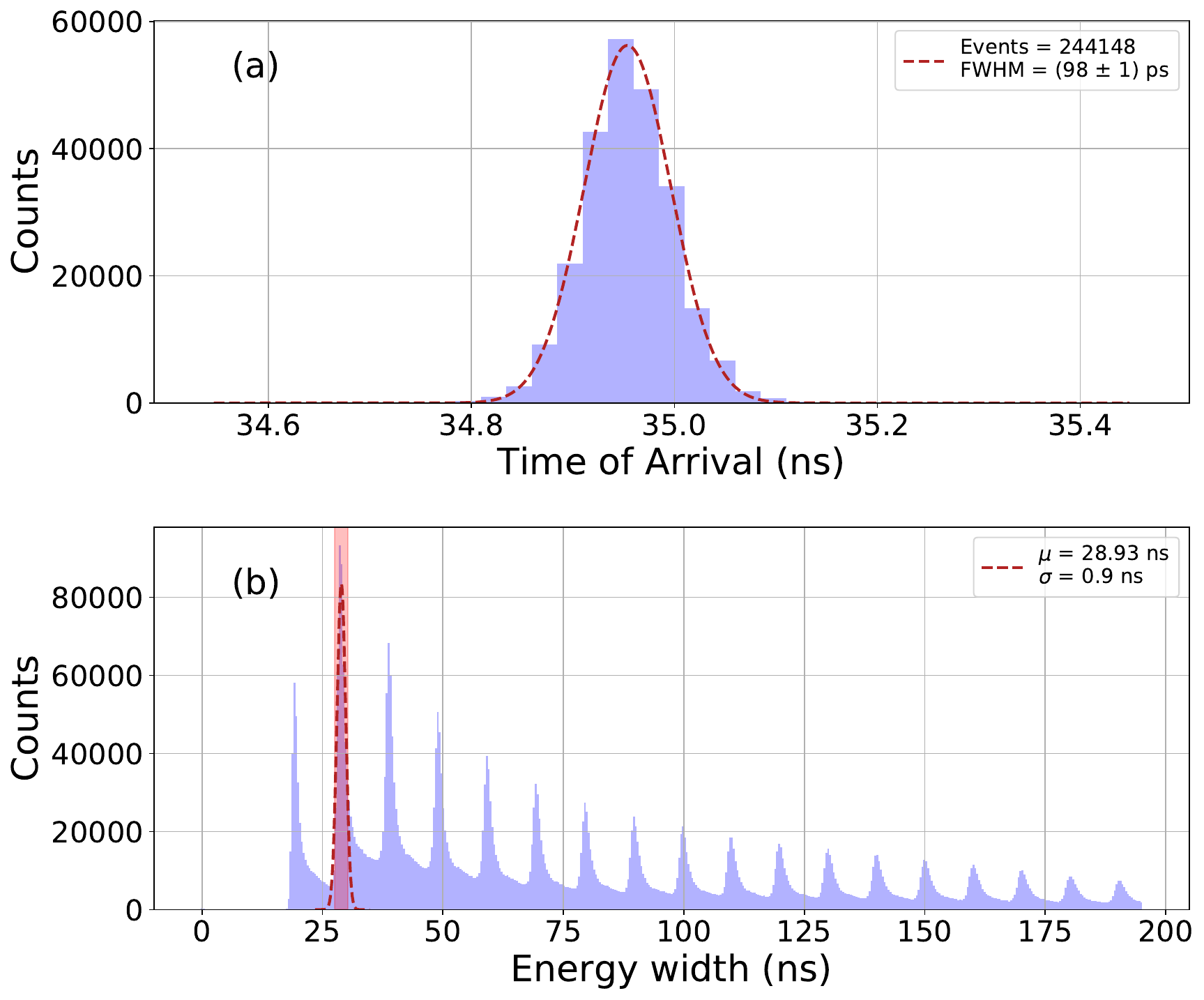}
        \caption{SPTR measurement with FastIC+ digital mode for FBK NUV-HD-MT LFv2 M0 3$\times$3 mm$^2$ at 16V of overvoltage.}
        \label{fig:SPTR_TDC}
    \end{minipage}
    \hfill
    \begin{minipage}[t]{0.49\textwidth}
        \centering
        \includegraphics[width=\textwidth]{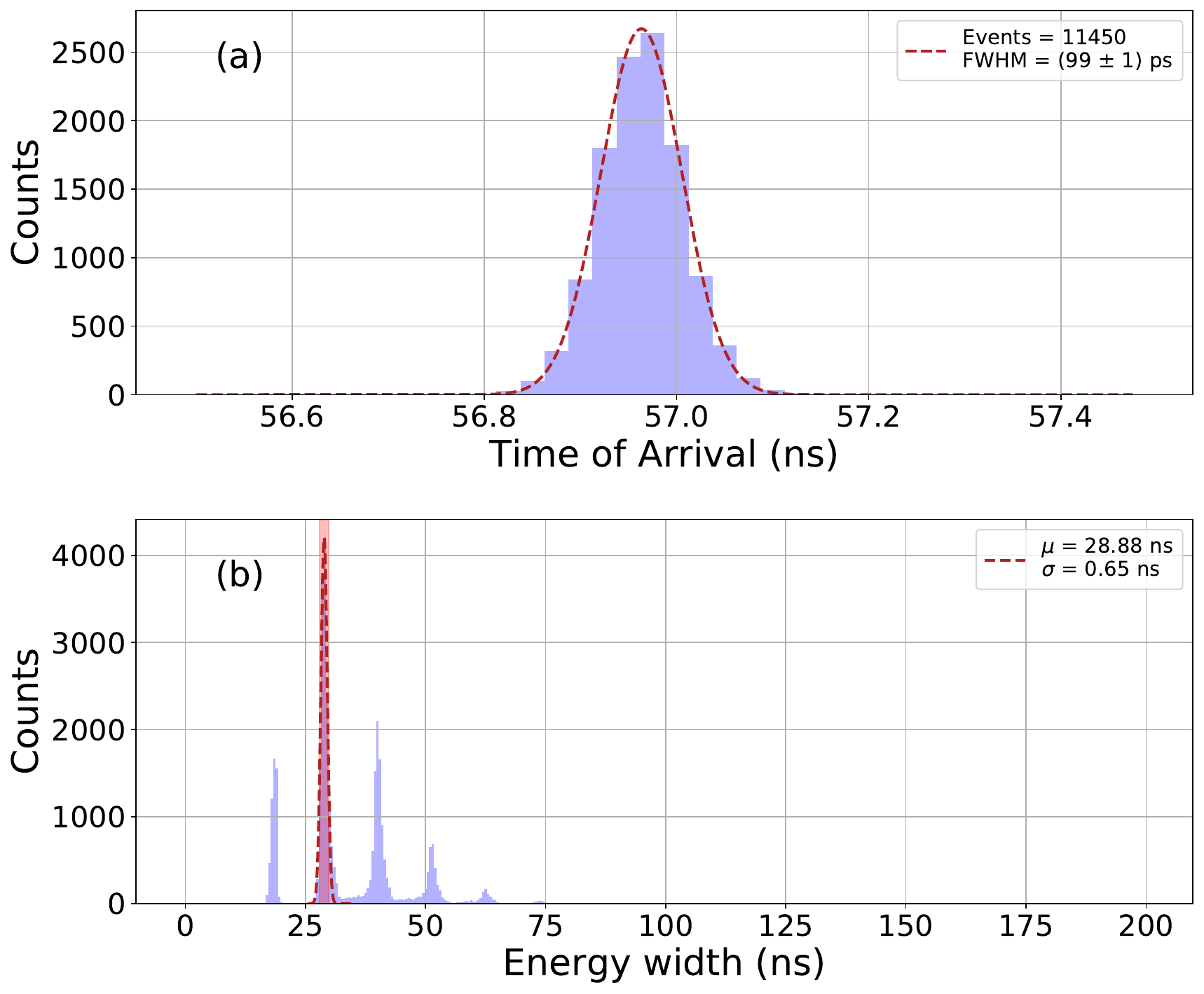}
        \caption{SPTR measurement with FastIC+ analog mode for FBK NUV-HD-MT LFv2 M0 3$\times$3 mm$^2$ at 16V of overvoltage.}
        \label{fig:SPTR_Scope}
    \end{minipage}
\end{figure*}

Figs.~\ref{fig:SPTR_TDC} and~\ref{fig:SPTR_Scope} compares the results obtained in digital and analog mode, respectively. To obtain these plots we used an FBK NUV-HD-MT LFv2 M0 SiPM (3$\times$3 mm$^2$, see Table~\ref{tab:SiPMs}) and operated it at an overvoltage of 16 V. The lower plots (Fig.~\ref{fig:SPTR_TDC}b and Fig.~\ref{fig:SPTR_Scope}b) show the Energy distributions. We selected 1-phe events (denoted with a red area in the Energy plot) to build the ToA distributions of Fig.~\ref{fig:SPTR_TDC}a and Fig.~\ref{fig:SPTR_Scope}a to extract the SPTR. Both measurements yielded a similar SPTR value, with (98 $\pm$ 1) ps and (99 $\pm$ 1) ps for the digital and analog measurements, respectively. Therefore, the TDC does not degrade the time resolution compared to the analog results, indicating that the contribution of the jitter of the TDC is not significant. 

Furthermore, Fig.~\ref{fig:SPTR_TDC}b shows that the energy readout has sufficient SNR to identify the different photons. Note that the distribution in Fig.~\ref{fig:SPTR_TDC}b is not Poissonian. This is because it combines acquisitions at different light intensities, which were taken to show FastIC+ capability to resolve events higher than 10 phe to evaluate the system's jitter of FastIC+ TDC (see Sec.~\ref{sec:system_jitter_TDC}).

\subsection{System's jitter FastIC+ TDC}
\label{sec:system_jitter_TDC}

\begin{figure}[htpb!]	
    \includegraphics[width=\columnwidth]{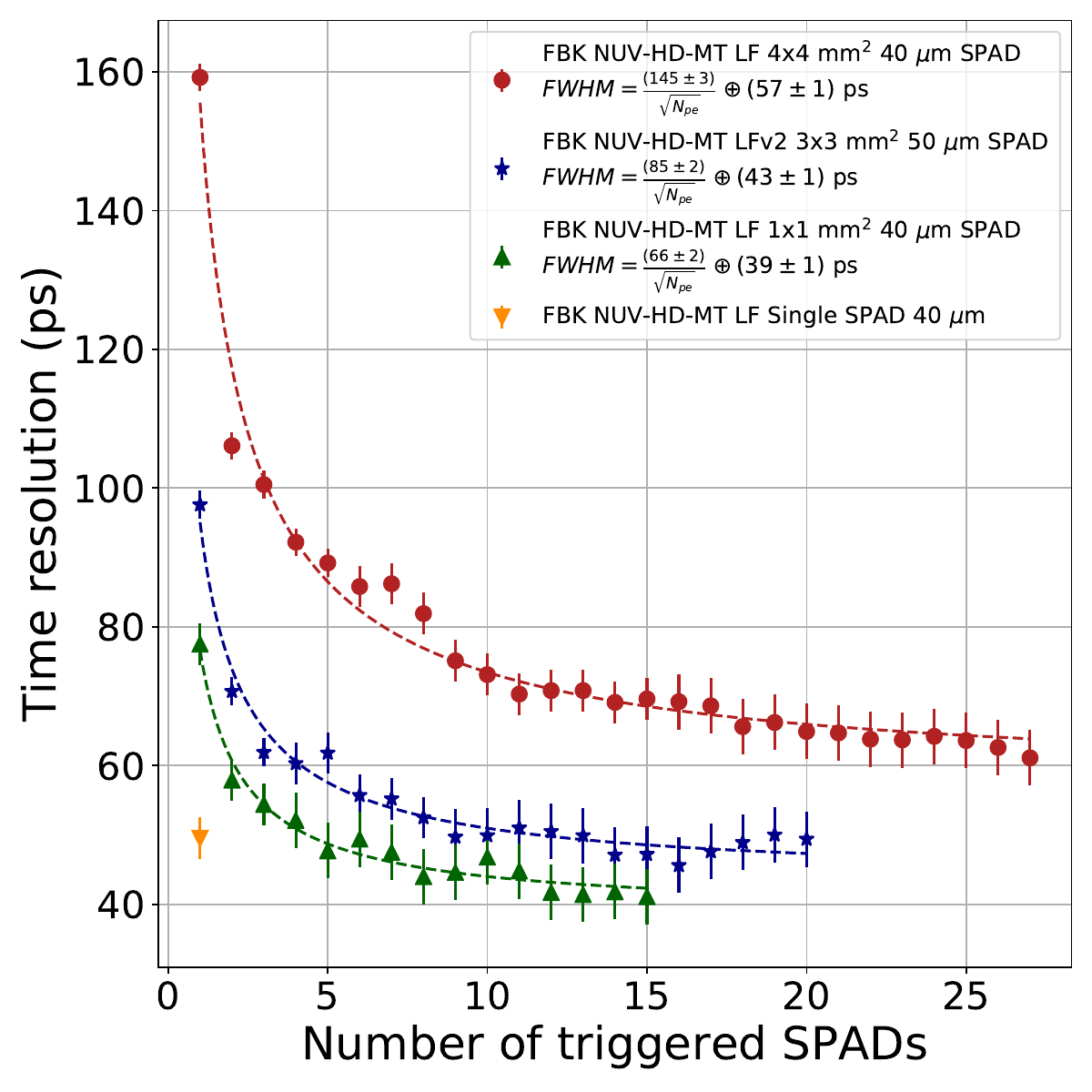}
    \caption{Time resolution for different number of fired SPADs obtained in digital mode for different SiPMs. All measurements were taken at 16V of overvoltage. Data was fitted to Eq.~\ref{eq:TR-NPTR}. }
    \label{fig:SPTR-system-jitter}
\end{figure}

Fig.~\ref{fig:SPTR-system-jitter} shows the time resolution of the FastIC+ TDC for different numbers of fired SPADs. To build this plot, we constructed the ToA distributions and computed the time resolution ($\Delta T$) for each photoelectron peak (i.e. the different peaks observed in Fig.~\ref{fig:SPTR_TDC}b), as described in Sec.~\ref{sec:Res-TDC_vs_Scope}. As a first-order approximation, the time resolution for $N_{\text{p.e.}}$ photons can be estimated by fitting the data to Eq.~\ref{eq:TR-NPTR}, as detailed in \cite{CHYTKA2019,Gundacker2020a}, 

\begin{equation}
    \Delta T (N_{p.e.})= 2 \sqrt{2 \; ln(2)} \cdot \sqrt{\frac{\sigma^{
    2}_{PJ}}{N_{p.e.}} + \sigma^{2}_{syst}} \; ,
    \label{eq:TR-NPTR}
\end{equation}

\noindent where $\sigma_{PJ}$ accounts for the single-photon electronic jitter and the intrinsic SPTR of the sensor. This parameter directly depends on the number of photons. The second parameter, $\sigma_{syst.}$, represents the constant jitter contribution of the system, which is independent of the number of photons. This term comprises the laser pulse shape, the laser trigger jitter, the electronic jitter at the maximum slew rate achievable by the electronics and the acquisition jitter.

The time resolution improves as the number of triggered SPADs increases. This improvement is mainly due to two factors. First, when multiple SPADs are triggered, the timing jitter associated with each individual detection (i.e., the intrinsic SPTR of the sensor) is averaged, resulting in a more accurate estimate of the photon's arrival time. Second, the combined signal from multiple detections leads to a higher slew rate in the analog output, thereby reducing the electronic jitter contribution to the overall jitter.

The observed timing differences between sensors can be attributed to the sensors themselves. Since all measurements were performed under identical conditions (using the same laser setup and the same TDC), the time resolution differences can be attributed to the SiPM. In particular, smaller sensors (e.g., 1$\times$1 mm$^2$ SiPMs) exhibit a lower jitter compared to larger ones. This is due to reduced electronic noise and optical noise associated with a smaller sensor area.

\subsection{Digital vs analog performance for CTR}

\begin{figure*}[htb!]
    \centering
    \begin{minipage}[t]{0.49\textwidth}
        \centering
        \includegraphics[width=\textwidth]{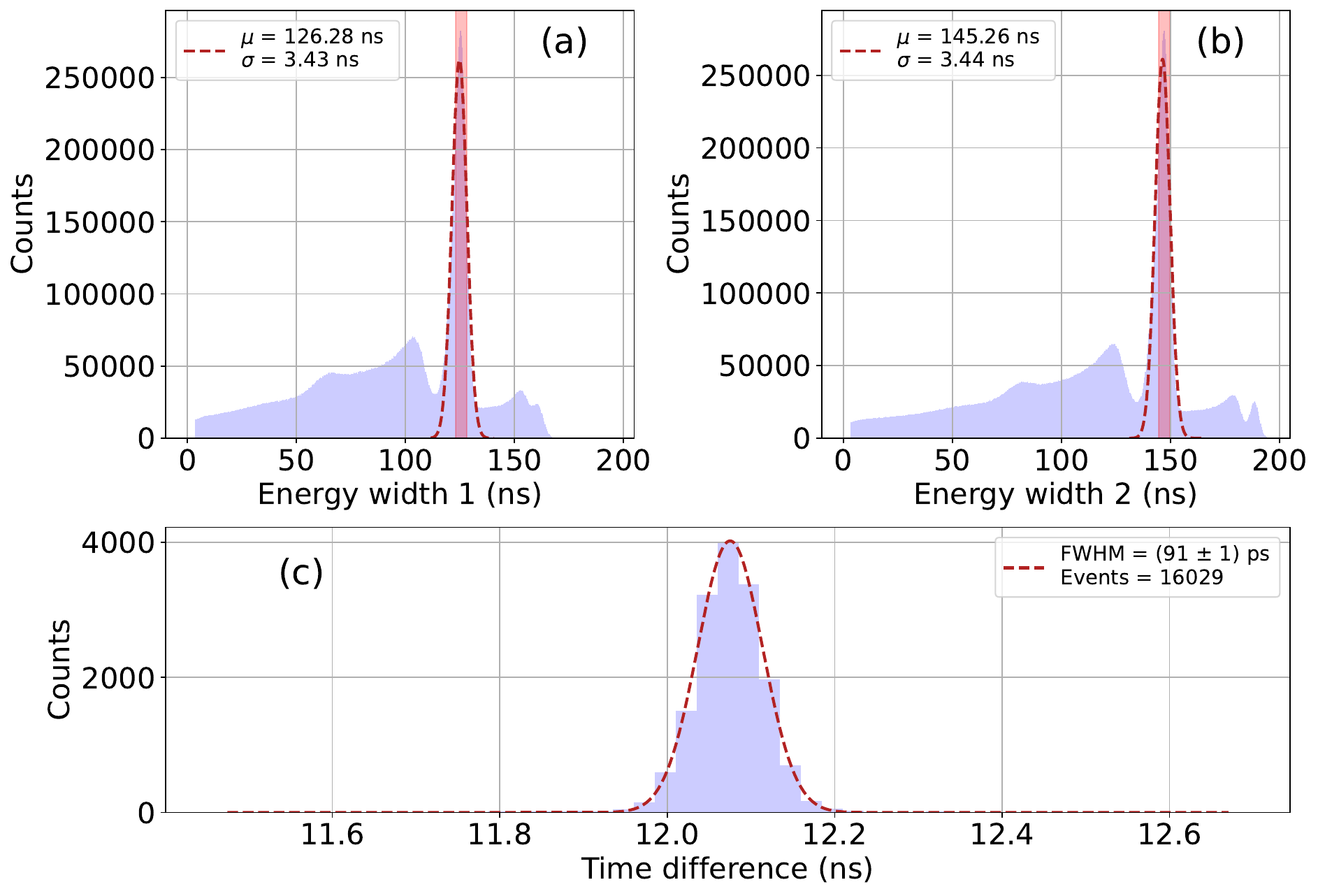}
        \caption{CTR measurement with FastIC+ digital mode for FBK NUV-HD-MT LFv2 M0 3$\times$3 mm$^2$ SiPMs coupled to 2$\times$2$\times$3 mm$^3$ LYSO:Ce,Ca crystals from TAC at 10 V of overvoltage.}
        \label{fig:CTR_TDC}
    \end{minipage}
    \hfill
    \begin{minipage}[t]{0.49\textwidth}
        \centering
        \includegraphics[width=\textwidth]{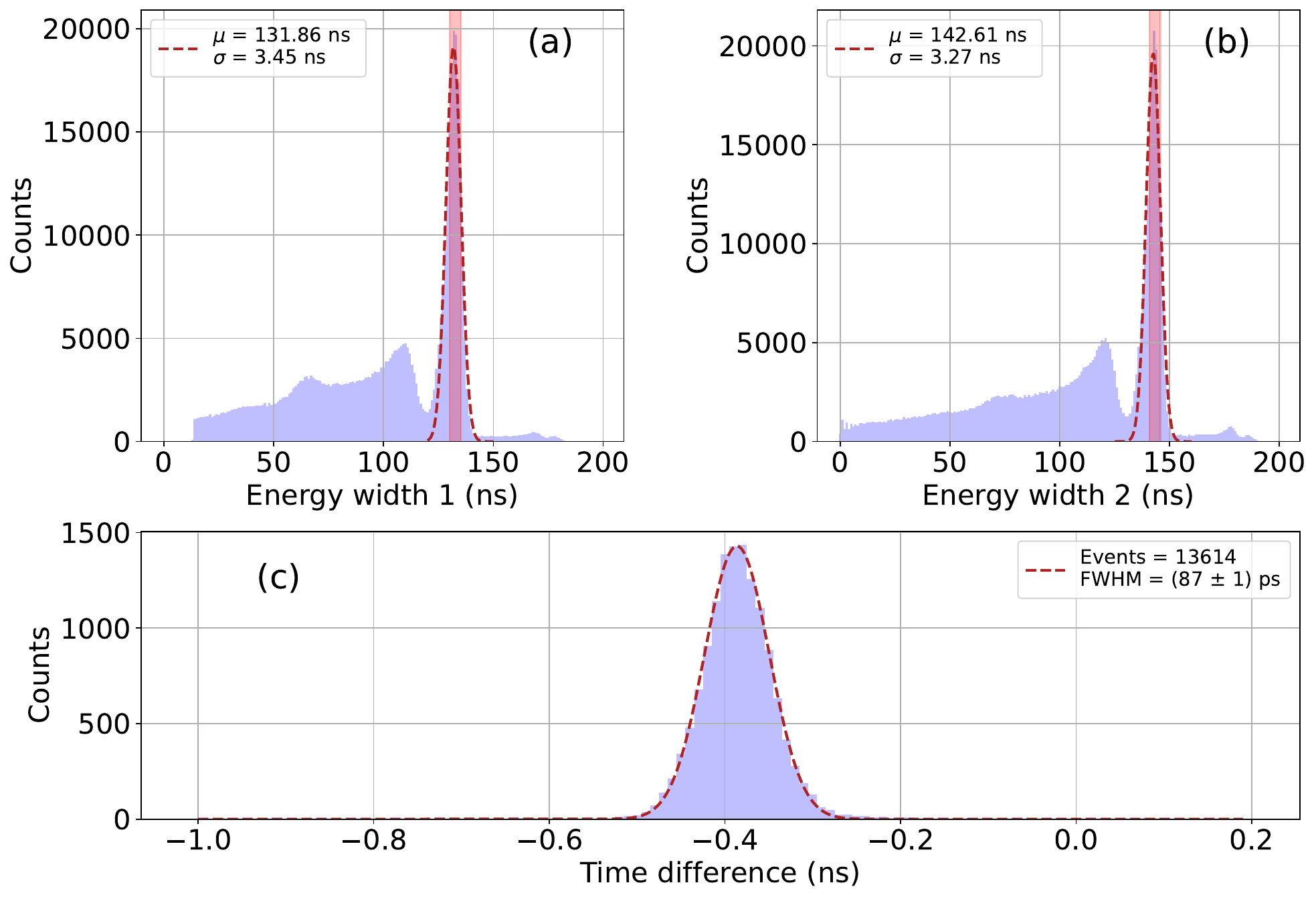}
        \caption{CTR measurement with FastIC+ analog mode for FBK NUV-HD-MT LFv2 M0 3$\times$3 mm$^2$ SiPMs coupled to 2x2x3 mm$^3$ LYSO:Ce,Ca crystals from TAC at 10 V of overvoltage.}
        \label{fig:CTR_Scope}
    \end{minipage}
\end{figure*}

Figs.~\ref{fig:CTR_TDC} and ~\ref{fig:CTR_Scope} compare the FastIC+ CTR measured with the digital and analog mode, respectively. Those plots were obtained using FBK NUV-HD-MT LFv2 3$\times$3 mm$^2$ SiPMs coupled to LYSO:Ce,Ca crystals of 2$\times$2$\times$3 mm$^3$ from TAC. Figs~\ref{fig:CTR_TDC}a, ~\ref{fig:CTR_TDC}b, ~\ref{fig:CTR_Scope}a and ~\ref{fig:CTR_Scope}b show the $^{22}$Na energy spectra recorded with the two detectors in both modes. As seen in the figures, the photopeaks are centered at similar energies when comparing analog and digital signal for both detectors. Figs~\ref{fig:CTR_TDC}c and ~\ref{fig:CTR_Scope}c show the difference in the arrival time of the two detectors for 511 keV events (denoted as the red regions in the energy plots). Both measurements yielded a similar CTR value of (91 $\pm$ 1) ps and (87 $\pm$ 1) ps in digital and analog mode, respectively, achieving a sub-100 ps time resolution. As observed in Sec.~\ref{sec:Res-TDC_vs_Scope} for low-light-level signals, the TDC contribution to the time resolution is minimal, and only a small difference is observed when comparing with the analog output for large signals.

\subsection{CTR performance using different scintillators}

\begin{figure*}[htb!]
    \begin{minipage}[t]{0.49\textwidth}
        \centering
        \includegraphics[width=\columnwidth]{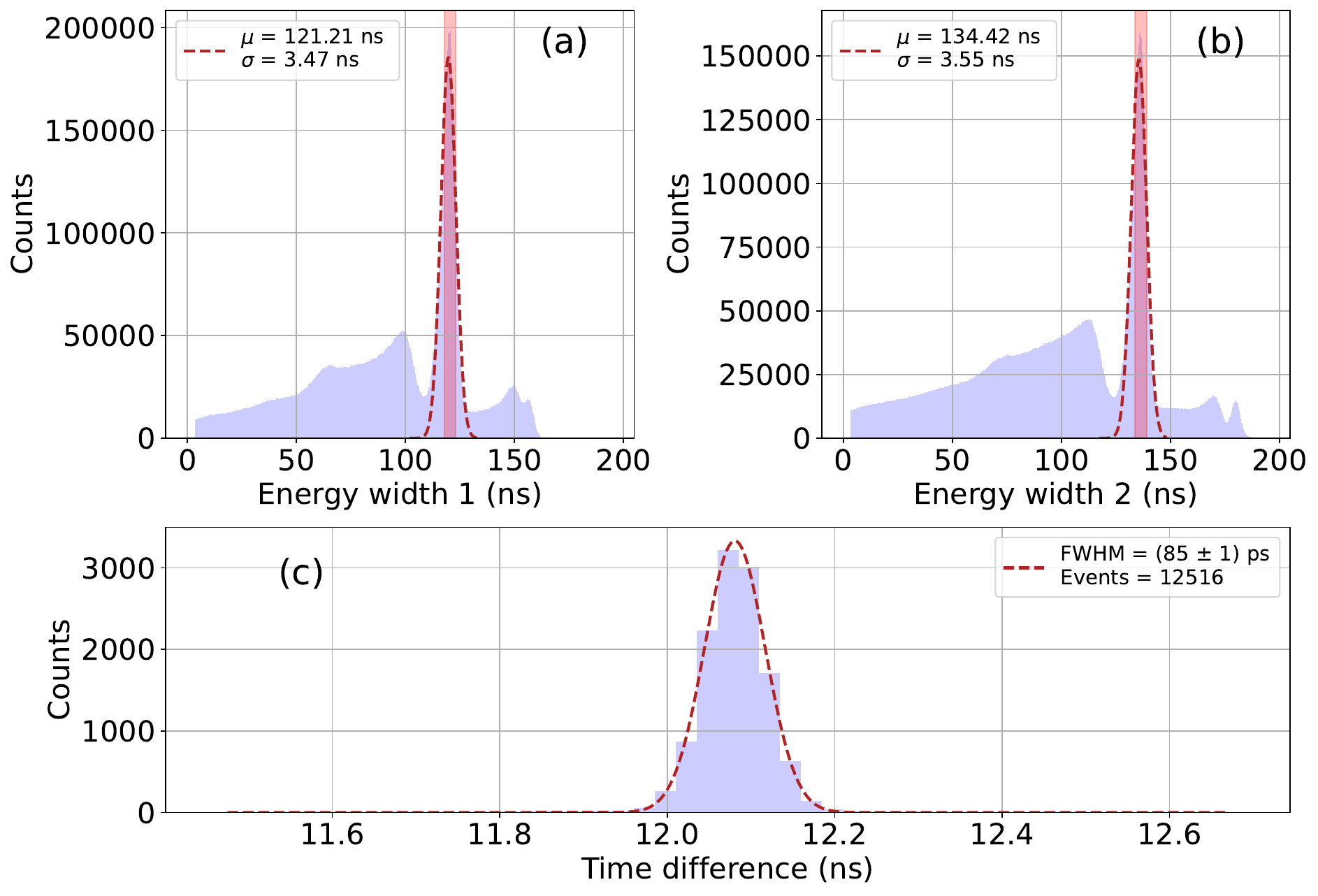}
        \caption{CTR measurement with FastIC+ TDC for FBK NUV-HD-MT LFv2 M0 3$\times$3 mm$^2$ SiPMs coupled to Fast-LGSO 2$\times$2$\times$3 mm$^3$ crystals from Oxide at 10 V of overvoltage.}
        \label{fig:CTR_Oxide_2x2x3}
    \end{minipage}
    \hfill
    \begin{minipage}[t]{0.49\textwidth}
        \centering
        \includegraphics[width=\columnwidth]{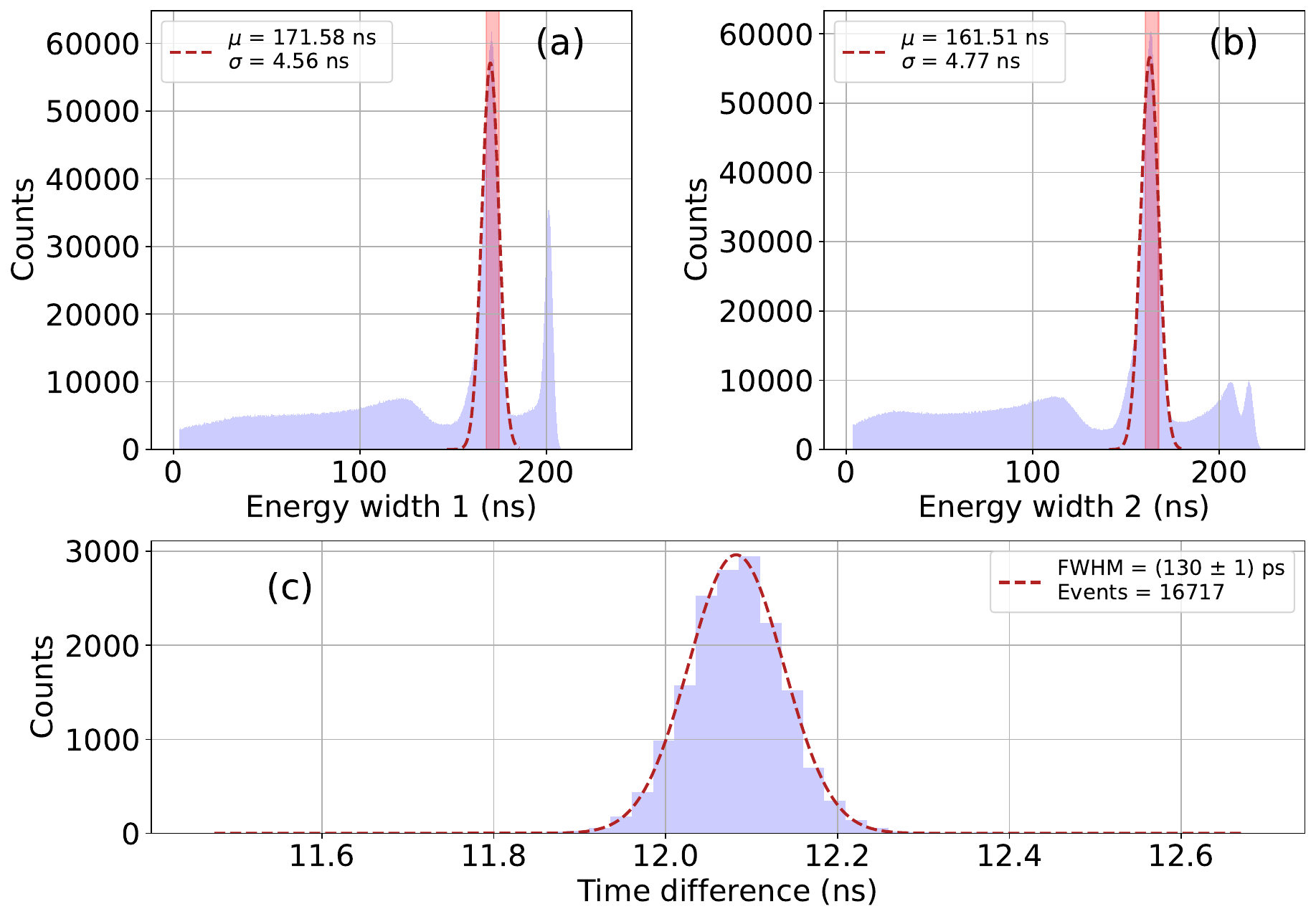}
        \caption{CTR measurement with FastIC+ TDC for FBK NUV-HD-MT LF M0 4$\times$4 mm$^2$ SiPMs coupled to LYSO:Ce,Ca 2.8$\times$2.8$\times$20 mm$^3$ crystals from TAC at 10 V of overvoltage. The peak around $\sim$ 200 ns observed in (a) is due to signal saturation at the configured energy gain in this readout channel. }
        \label{fig:CTR_TAC_2.8x2.8x20}
    \end{minipage}
\end{figure*}

Table~\ref{tab:CTR-results} presents the best CTR results achieved for a different combination of scintillators and SiPMs using the FastIC+ TDC. The lowest timing resolution achieved was (85 $\pm$ 1) ps for Fast-LGSO crystals from Oxide, coupled to FBK NUV-HD-MT LFv2 M0 SiPMs with a 3$\times$3 mm$^2$ of active area, as seen in Fig.~\ref{fig:CTR_Oxide_2x2x3}. We also tested 2$\times$3$\times$3 mm$^3$ crystals from different manufacturers using the same SiPMs, obtaining sub-100 ps results for all of them. Additionally, we evaluated the same LYSO:Ce,Ca crystals from TAC with different SiPMs to study the contribution of the photosensor. For SiPMs based on the same technology but with different active areas and different SPAD size, i.e., FBK 3$\times$3 mm$^2$ and 50~$\mu$m and FBK 4$\times$4 mm$^2$ and 40~$\mu$m, we observed a 8 ps degradation in the timing resolution. However, the most significant difference was found when comparing HPK and FBK sensors using the same TAC crystal, with a $\sim$ 20 ps deterioration when using the HPK.

We also evaluated the CTR performance of the FastIC+ TDC for longer crystals. The best result for 20 mm crystals was (130 $\pm$ 1) ps, achieved with a LYSO:Ce,Ca crystal from TAC measuring 2.8$\times$2.8$\times$20 mm$^3$, coupled to an FBK NUV-HD-MT LF 4$\times$4 mm$^2$ SiPM, as shown in Fig.~\ref{fig:CTR_TAC_2.8x2.8x20}.

Large crystals exhibited similar CTR values ranging from 130 to 140 ps. A degradation of $\sim$ 7 ps is observed when increasing the area of the crystal from 2.8$\times$2.8 mm$^2$ to 3$\times$3$^2$ for the TAC manufacturer. For Saint-Gobain and CPI crystals, both measurements provided a similar CTR value of (134 $\pm$ 1) ps and (135 $\pm$ 1) ps, respectively, despite the larger area of the CPI crystal. All 20 mm crystals were coupled to the same 4$\times$4 mm$^2$ SiPMs, which have a larger active area than the crystals to minimize light losses due to eventual misalignments.

% We also evaluated the CTR performance of the FastIC+ TDC for longer crystals. Fig.~\ref{fig:CTR_CPI_3.12x3.12x10} presents a CTR value of 121 $\pm$ 1 ps for a LYSO:Ce:0.2$\%$:Ca:0.2$\%$ 3.2x3.2x10 mm$^3$ crystal coupled to FBK NUV-HD-MT LF 4x4 mm$^2$ SiPM. This result is in agreement with the CTR value of 117 $\pm$ 4 ps reported with the analog FastIC with the same crystal coupled to FBK NUV-HD-MT LFv2 3.2x3.12 mm$^2$ SiPM \cite{AntonioFastIC2024}.

% \begin{figure}[htp]	
%     \includegraphics[width=\columnwidth]{Figures/results/CTR-CPI-3.12x3.12x10.pdf}
%     \caption{CTR measurement with FastIC+ TDC for FBK NUV-HD-MT LF 4x4 mm$^2$ coupled to a LYSO:Ce:0.2$\%$:Ca:02$\%$ 3.12x3.12x10 mm$^3$ crystal at 10 V of overvoltage.}
%     \label{fig:CTR_CPI_3.12x3.12x10}
% \end{figure}

\begin{table*}[htp!]
    \centering
    \caption{Best CTR results performed with FastIC+ TDC.}
    \begin{tabular}{|c|c|c|c|}
    \hline
    \rowcolor[HTML]{C0C0C0}
    {\color[HTML]{000000} \textbf{Crystal}} & {\color[HTML]{000000} \textbf{Size (mm$^3$)}} & {\color[HTML]{000000} \textbf{SiPM (mm$^2$)}} & {\color[HTML]{000000} \textbf{CTR ($\pm$ 1 ps FWHM)}} \\
    \hline
    \rowcolor[HTML]{E3EBAB}
    Fast-LGSO (Oxide) & 2$\times$2$\times$3 & FBK NUV-HD-MT LFv2 M0 3$\times$3 & 85 \\
    \hline
    \rowcolor[HTML]{D9EBF1}
    LYSO:Ce,Ca (CPI) & 2$\times$2$\times$3 & FBK NUV-HD-MT LFv2 M0 3$\times$3 & 92 \\
    \hline
    \rowcolor[HTML]{E3EBAB}
    LYSO:Ce,Ca (TAC) & 2$\times$2$\times$3 & FBK NUV-HD-MT LFv2 M0 3$\times$3 & 91 \\
    \hline
    \rowcolor[HTML]{D9EBF1}
    LYSO:Ce,Ca (TAC) & 2$\times$2$\times$3 & FBK NUV-HD-MT LF M0 4$\times$4 & 99 \\ 
    \hline
    \rowcolor[HTML]{E3EBAB}
    LYSO:Ce,Ca (TAC) & 2$\times$2$\times$3 & HPK S13360-3050VE 3$\times$3 & 110 \\
    \hline
    \rowcolor[HTML]{D9EBF1}
    LYSO:Ce,Ca (TAC) & 2.8$\times$2.8$\times$20 & FBK NUV-HD-MT LF M0 4$\times$4 & 130 \\
    \hline
    \rowcolor[HTML]{E3EBAB}
    LYSO:Ce,Ca (TAC) & 3$\times$3$\times$20 & FBK NUV-HD-MT LF M0 4$\times$4 & 137 \\ 
    \hline
    \rowcolor[HTML]{D9EBF1}
    LYSO:Ce:Ca (Saint-Gobain) & 3$\times$3$\times$20 & FBK NUV-HD-MT LF M0 4$\times$4 & 134 \\ 
    \hline
    \rowcolor[HTML]{E3EBAB} 
    LYSO:Ce:Ca (CPI) & 3.12$\times$3.12$\times$20 & FBK NUV-HD-MT LF M0 4$\times$4 & 135 \\ 
    \hline
    \end{tabular}
    \label{tab:CTR-results}
\end{table*}

% The analog performance of the FastIC was evaluated in \cite{FastIC_TRPMS}. When coupled with FBK NUV-HD LF2 M0 SiPMs, the FastIC provided a Single Photon Time Resolution (SPTR) of (151 ± 3) ps FWHM and a Coincidence Time Resolution (CTR) of (127 ± 3) ps FWHM, when coupled to LYSO:Ce:0.2$\%$Ca of 3.13 $\times$ 3.13 $\times$ 20 mm$^3$. 

% We have done a preliminary study of the Single Shot Precision of the full TDC.
% The expected TDC resolution should be around 9 ps
% Assuming an Electronics Noise contribution for 1 pe of 30 ps, the Single Shot Precision of FastIC+ should be around 31.4 ps, an increase in 4.5\% wrt the previous FastIC ASIC.

% Regarding the Power Consumption, we expect to have a Power Consumption lower than 27 mW
% 3.4 mW/ch
% FastIC has 12 mW/ch of power consumption.
% In FastIC+ we can power down the individual SLVS binary outputs and save 2.75 mW/ch
% The expected power consumption will be around 12.65 mW/ch, an extra 5.4%.

    \section{Discussion} \label{sec:discussion}

The new FastIC+ ASIC has demonstrated excellent timing resolution while introducing an on-chip digitization with only a 5\% increase in power consumption compared to its predecessor, FastIC. This digitization, achieved through the internal TDC of the FastIC+ ASIC, does not introduce any significant jitter in the time resolution measurements, thereby simplifying data acquisition at the system level. Experimental results confirm that the timing resolution obtained from the analog output is comparable with the on-chip TDC results, both in SPTR and CTR measurements. Furthermore, the minimal contribution of the TDC can also be observed when analyzing the system's jitter. As shown in Fig.~\ref{fig:SPTR-system-jitter}, the differences in time resolution clearly depend on the size of the SiPM, demonstrating that the measurement is dominated by SiPM-related jitter rather than acquisition jitter, which is independent of sensor size.

The best CTR obtained using the FastIC+ TDC was (85 $\pm$ 1) ps for Fast-LGSO crystals from Oxide measuring 2$\times$2$\times$3 mm$^3$, coupled to FBK NUV-HD-MT LFv2 M0 SiPMs with a 3$\times$3 mm$^2$ sensitive area. The performance obtained with this crystal was few ps better compared to the other crystals of the same size. When comparing the performance of crystals from different manufacturers, we observe an improvement of 6 ps and 7 ps relative to TAC and CPI, respectively. This improvement might be related to the faster decay time of Fast-LGSO crystals, although there are many factor affecting the time resolution such as scintillator light yield, efficiency of the crystal-SiPM coupling, etc.

We also studied the impact of the SiPM area on CTR measurements. When the same LYSO:Ce,Ca crystal of size 2$\times$2$\times$3 mm$^3$ from TAC is coupled to a SiPM with a larger sensitive area and smaller pixel size, an 8~ps degradation is observed. This deterioration is attributed to the increase in SPTR, as the 4$\times$4 mm$^2$ SiPM presents a higher SPTR (approximately 60~ps worse) compared to the 3$\times$3~mm$^2$ SiPM, as shown in Fig.~\ref{fig:SPTR-system-jitter}.

For 20 mm crystals, all measurements provided similar CTR values for different manufacturers and crystal sizes. The 3.12$\times$3.12$\times$20 mm$^3$ crystals from CPI had been measured with FastIC using FBK NUV-HD LFv2 M0 3.12$\times$3.12 mm$^2$ SiPMs, reaching a CTR value of 127 ps with analog signals \cite{AntonioFastIC2024}. We observed a deterioration of $\sim$ 8 ps when comparing with FBK NUV-HD-MT LF M0 4$\times$4 mm$^2$, which can be attributed to a larger sensor area and smaller pixel size. 

Table \ref{tab:ASICS} presents a comparison of various ASICs and discrete electronics studied in the literature for ToF-PET applications. To the best of the author's knowledge, FastIC+ is the first ASIC with on-chip digitization, achieved through an internal TDC, to attain a sub-100 ps CTR for short crystals and sub-140 ps for 20 mm-long crystals. 

\begin{table*}[htp!]
    \centering
    \caption{Comparison of state-of-the-art electronics employed in TOF-PET applications. }
    \begin{tabular}{|c|c|c|c|c|c|c|}
    \hline
    \rowcolor[HTML]{C0C0C0} 
    {\color[HTML]{000000} \shortstack{ \textbf{ASIC} \\ \textbf{}}} & {\color[HTML]{000000} \shortstack{ \textbf{Power} \\ \textbf{(mW/ch)}} }  & {\color[HTML]{000000} \shortstack{ \textbf{Crystal} \\ \textbf{[mm$^3$]}} } & {\color[HTML]{000000} \shortstack{ \textbf{SiPM} \\ \textbf{[mm$^2$]}} } & {\color[HTML]{000000} \shortstack{ \textbf{CTR} \\ \textbf{(ps FWHM)} }  } & {\color[HTML]{000000} \shortstack{ \textbf{DAQ}$^a$ \\ \textbf{}} } & {\color[HTML]{000000} \shortstack{ \textbf{Reference} \\ \textbf{}} } \\
    \hline
    \rowcolor[HTML]{E3EBAB}
    NINO & 27 & LSO:Ce,0.4$\%$Ca [2$\times$2$\times$3] & FBK NUV-HD [3$\times$3] & 73 & Osc. & \cite{Gundacker_2019} \\ \hline
    \rowcolor[HTML]{D9EBF1}
    Radioroc & 3.3$^b$  & LYSO:Ce,Ca [2$\times$2$\times$3] & Broadcom NUV-MT [4$\times$4] & 83 & Osc. & \cite{saleem23,Weeroc} \\ 
    \hline
    \rowcolor[HTML]{E3EBAB} 
    Radioroc & 3.3$^b$ & LYSO:Ce,Ca [3$\times$3$\times$20] & Broadcom NUV-MT [4$\times$4] & 127 & Osc. & \cite{saleem23,Weeroc} \\ 
    \hline
    \rowcolor[HTML]{D9EBF1}
    TOFPET2c & 8 & LYSO:Ce,Ca [3$\times$3$\times$19] & Broadcom NUV-MT [3.8$\times$3.8] & 157 & \textbf{TDC} & \cite{Nadig_2023} \\ 
    \hline
    \rowcolor[HTML]{E3EBAB}
    HF readout & 143 & LYSO:Ce,Ca [2$\times$2$\times$3] & Broadcom NUV-MT [3.8$\times$3.8] & 56 & Osc. + IC & \cite{Nadig_2023} \\
    \hline
    \rowcolor[HTML]{D9EBF1}
    HF readout & 143 & LYSO:Ce,Ca [3$\times$3$\times$19] & Broadcom NUV-MT [3.8$\times$3.8] & 95 & Osc. + IC & \cite{Nadig_2023} \\
    \hline
    \rowcolor[HTML]{E3EBAB}
    FastIC & 12 & LSO:Ce:0.2$\%$:Ca [2$\times$2$\times$3] & FBK NUV-HD LFv2 M0 [3.12$\times$3.2]  & 76 & Osc. & \cite{AntonioFastIC2024} \\ 
    \hline
    \rowcolor[HTML]{D9EBF1}
    FastIC & 12 & LYSO:Ce,0.2$\%$:Ca [3.13$\times$3.13$\times$20] & FBK NUV-HD LFv2 M0 [3.12$\times$3.2] & 127 & Osc. & \cite{AntonioFastIC2024} \\ 
    \hline
    \rowcolor[HTML]{E3EBAB}
    FastIC+ & 12.5 & Fast-LGSO [2$\times$2$\times$3] & FBK NUV-HD-MT LFv2 M0 [3$\times$3] & 85 & \textbf{TDC} & This work \\ 
    \hline
    \rowcolor[HTML]{D9EBF1}
    FastIC+ & 12.5 & LYSO:Ce,Ca [2.8$\times$2.8$\times$20] & FBK NUV-HD-MT LF M0 [4$\times$4]  & 130 & \textbf{TDC} & This work \\
    \hline
    \end{tabular}
    \label{tab:ASICS}

    \begin{tablenotes}
    \item $^a$ Data Acquisition Method. All ASICs include an internal comparator to obtain the timestamps, which outputs a binary signal that can be acquired either by an oscilloscope (Osc.) or by a TDC. Notice that HF readout outputs analog signals. Therefore, an ideal comparator (denoted as IC) is applied to data to obtain the CTR.\\
    \item $^b$ Power consumption of Radioroc2.
    \end{tablenotes}

\end{table*}

Although FastIC+ has demonstrated excellent timing performance, its CTR results are still higher than those compared to high-power discrete electronics, indicating room for improvement. As shown in Table~\ref{tab:ASICS}, the best results have been achieved with HF readouts. These systems provide lower electronic noise ($\sigma_{EN}$) compared to FastIC+, therefore reducing the jitter contribution, since it scales with $\sigma_{EN}/SR$, where $SR$ denotes the slew rate of the signal. Additionally, HF systems provide analog signals and they rely on fast oscilloscopes combined with ideal comparators to extract the ToA using an offline leading-edge comparator algorithm applied to the analog data. In contrast, FastIC+ employs an integrated on-chip comparator which introduces additional jitter to the measurement, degrading the timing measurement. However, the power consumption of HF readouts makes them unsuitable for large-scale systems where hundreds of detectors are involved. 

In terms of power consumption, FastIC+ has a higher power consumption than the TOFPET2c ASIC. It is important to note that the FastIC+ is a multipurpose ASIC designed to process not only SiPM signals but also those from MCPs or PMTs. These two types of sensors require a larger bandwidth for energy readout due to the short duration of their pulses. Therefore, the power consumption of the energy readout could be reduced if the FastIC energy readout were optimized specifically for SiPMs in PET applications.

The performance of FastIC with other types of scintillators, such as Cherenkov radiators and BGO crystals, has already been demonstrated in \cite{FastIC_PMB,AntonioFastIC2024,PintoFastic2015}. For instance, for BGO measurements it has already been show that time-walk correction improves CTR \cite{FastIC_PMB}. A well-known method involves measuring the rise time of the input signal to better identify the events contributing to the time of arrival. FastIC+ can provide a second timestamp to measure the rise time using the trigger signal. Observe that the trigger signal is generated as a fast-OR, meaning that only the timestamp of the fastest channel in the ASIC is available. Nevertheless, measuring the fastest channel may still provide sufficient information for CTR measurements with pixelated crystals.

FastIC+ capabilities for ToF-PET detectors can also be extended to dual-ended readout systems. These systems account for time-based DOI corrections and have already been implemented using high power consumption electronics, improving timing resolution not only for BGO crystals but also for LYSO \cite{KratochwilDualSide2025}. However, this approach comes at the cost of doubling the number of readout channels and increasing the mechanical complexity of the system.

	\section{Conclusions} \label{sec:Conclusions}
FastIC+ is a versatile read-out ASIC with internal digitization that exhibits low power consumption of 12.5 mW per channel, while achieving remarkable timing resolution. This ASIC is capable of measuring both positive and negative polarity sensors, providing digitized outputs for photon arrival time and energy. The internal TDC does not degrade the time resolution, as the measurements performed in digital and analog modes showed similar results.

A CTR of (85 $\pm$ 1) ps was achieved with FastIC+ when coupling Fast-LGSO crystals measuring 2$\times$2$\times$3 mm$^3$ with FBK NUV-HD-MT-LFv2 M0 3$\times$3 mm$^2$ SiPMs. For larger crystals, a CTR of (130 $\pm$ 1) ps was obtained using LYSO:Ce,Ca crystals measuring 2.8$\times$2.8$\times$20 mm$^3$ with FBK NUV-HD-MT LF M0 4$\times$4 mm$^2$ SiPMs. Furthermore, different 20 mm long crystal sizes from various manufacturers were tested, with all of them achieving sub-140 ps CTR results when coupled to the same 4$\times$4 SiPMs.

Future avenues for FastIC+ include implementing a double-ended readout system when applied to long crystals, allowing for additional timing corrections \cite{KratochwilDualSide2025}. Using these systems, it becomes possible to better identify the interaction point of the gamma in the crystal and apply depth-of-interaction corrections. Additionally, the use of BGO crystals for CTR measurements can also be explored. The combination of time and trigger comparators can help to determine the signal’s rise time and apply time walk corrections. Furthermore, FastIC+ is capable to digitize up to 8 different channels, making it versatile for working with arrays of SiPMs and enabling the simultaneous detection of multiple signals.

Thanks to the integration of the TDC in the chip, FastIC+ does not need external mechanisms like an FPGA to digitize the signals. This turns FastIC+ into a suitable chip to be employed in a complex system like a PET scanner.

% Future avenues for FastIC+ include the use of BGO crystals for CTR measurements. The combination of time and trigger comparators can help to determine the signal's rise time and apply time walk corrections. Additionally, further timing corrections can be explored for FastIC+ when applied to long crystals by implementing double-ended readout systems\cite{KratochwilDualSide2025}. Using these systems, it becomes possible to better identify the interaction point of the gamma in the crystal and apply depth-of-interaction corrections. Furthermore, FastIC+ is capable to digitize up to 8 different channels, making it versatile for working with arrays of SiPMs and enabling the simultaneous detection of multiple signals.
    \section{Acknowledgments} \label{sec:acknowledgments}

This work was supported by HORIZON EUROPE European Research Council,
Grant/Award Number: 101099896.
This study was also supported by MICIIN with funding from the European
Union NextGenerationEU fund (PRTR-C17.I1/AEI/10.13039/501100011033/Unión Europea NextGenerationEU/PRTR) and by the Generalitat de
Catalunya.

\bibliographystyle{IEEEtran}
%% Cambiar PATH!!!!!!!!!!!!!!!!!!!!!!!!!!!
\interlinepenalty=10000 
\bibliography{main}{} % bibliography data 

\end{document}